# The foundation of “Saint Véran – Paul Felenbok” astronomical observatory

J.-M. Malherbe, Observatoire de Paris, PSL Research University, LESIA, Meudon, France

Email : Jean-Marie.Malherbe@obspm.fr Date: 24 July 2024

ORCID identification : https://orcid.org/0000-0002-4180-3729

## ABSTRACT

This paper is dedicated to the memory of Paul Felenbok (1936-2020) who was astronomer at Paris-Meudon observatory, and founded in 1974, fifty years ago, a high altitude station (2930 m), above Saint Véran village in the southern Alps (Queyras). It was initially devoted to the study of the solar corona. Following solar eclipses (1970, 1973) observed with the Lallemand electronic camera, the main goal was to detect with this sensitive detector the structures of the far and hot corona in forbidden lines, using either narrow bandpass filters or spectroscopy. But everything had to be done prior to observations: a track, a house for astronomers, a dome and a complex instrument. We summarize here this fantastic adventure, which was partly successful in terms of scientific results and had to stop in 1982; however, the activity of the station resumed after 1989 under the auspices of the “AstroQueyras” association, which replaced the coronagraph by a 62 cm night telescope from Haute Provence observatory; the station extended later with two 50 cm telescopes, was rebuilt in 2015 and received the visit of thousands of amateurs.

## INTRODUCTION

After the total solar eclipse of Mauritania (Rösch, 1973, the famous eclipse chased during 74 minutes by the Concorde 001 prototype flying at 2200 km/h), the "Electronic Camera" group of Paul Felenbok (1936-2020, figure 1) of the Paris-Meudon Observatory, wanted to resume the observations of the solar corona with a Lallemand-type electronic camera mounted on a coronagraph. This layer is made of a hot ($10^6$ K – $10^7$ K) and tenuous plasma, more than $10^6$ times darker than the solar disk. For that reason, it is difficult to observe outside eclipses, so that Lyot (1932) developed a special instrument at Meudon (the coronagraph) in order to eliminate as best as possible diffused light by the lenses, which is brighter than the corona. In addition, high altitude stations are required to reduce the atmospheric diffusion, which is also brighter than the corona at see level (for that reason Lyot observed at Pic du Midi). Hence, Felenbok’s team had to find a suitable place and build an instrument for a fairly long period.

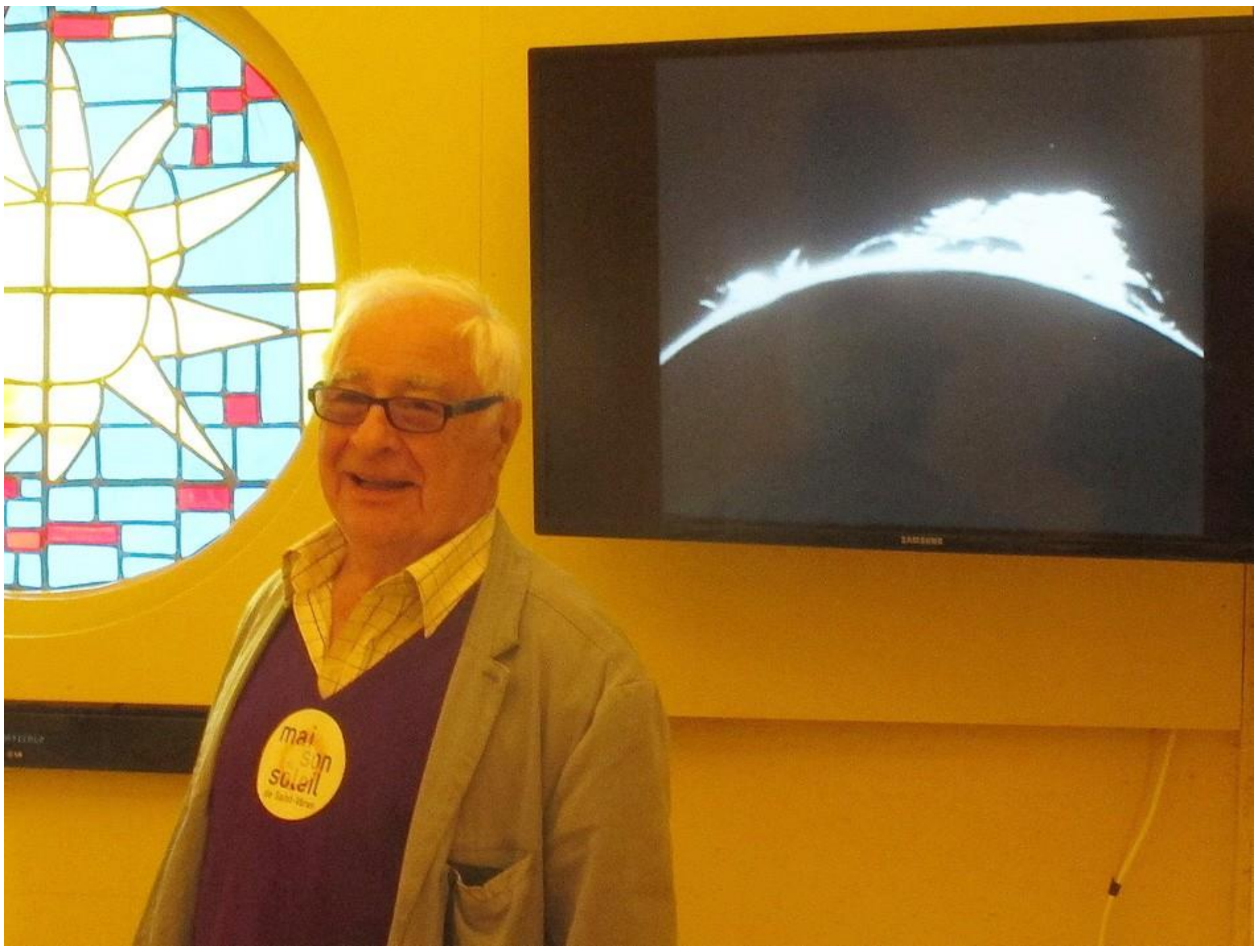

*Figure 1: Paul Felenbok (1936-2020) at the Sun’s house of Saint Véran in 2016 (courtesy D. Menel)*

The Pic du Midi in French Pyrénées (2870 m) was saturated at that time (no free coronagraph, no possibility of adding a heavy instrument to an existing mount), so that the group looked elsewhere. The preparation of the 1973 total eclipse took place in Auron (southern Alps), at the top of the cable car, an easily accessible location. However, the National Institute of Astronomy and Geophysics (INAG) of the CNRS (now INSU), organized a campaign of sites in 1970 and 1971, and identified the “Pic de Château Renard” (above Saint Véran) as the best night site in France (but for solar observations, it is always worse because of the heating of the ground). Indeed, INAG was starting a project for a large night telescope of the 4-metre class, which finally became operational in 1979 in Hawaii in collaboration with Canada and Hawaii University, at an altitude of 4200 m (Mauna Kea). It is the CFHT, whose dome is so large that it could house a 11-m multi-mirror telescope (in project). But let's go back to 1974: INAG recommended the development of the coronagraph in Saint Véran, in order to start the development of a new site for future small or medium size instruments, and to characterize the seeing for long term observations.

## THE 7-METRE DOME OF PARIS OBSERVATORY BEFORE TRANSFER TO SAINT VÉRAN

The budget for Saint Véran was rather limited, because it was not a national project, therefore Paris observatory accepted to dismount the unused 7-m dome located on the roof of the “Perrault” building. This dome has a long story, which begins in 1858 when Urbain Le Verrier (1811-1877) decided to put an instrument on the west tower. A 7-m wooden dome was ordered to Joseph Jean, a carpentry manufacturer, in May 1857 and delivered in 1858. Le Verrier equipped it with an equatorial supporting a 30 cm refractor (5.25 m focal length). This instrument was built under the supervision of the astronomer Villarceau (1813-1883), by Eichens, who owned the Secrétan workshops. It was delivered in March 1858, and the first observation took place in October 1858. Villarceau considered that the objective was not of sufficient quality; it was corrected and commissioned in November 1860. The mechanics was excellent, but the optics was not so good. A contract was signed with the optician Martin (1824-1896) for the supply of a new lens, delivered in 1878. It was not installed until 1884 when Bigourdan (1851-1932) noticed systematic errors introduced by the original lens. During maintenance work, a fire started in July 1904, but it was quickly brought under control by the firefighters and a repair was undertaken. The dome was repainted again in 1931. From 1933 to 1949, the medical doctor Paul Baize (1901-1995), a volunteer collaborator, carried out measurements of double stars. He wrote: "After the war, the dome, pierced in many places by flak, began to let the rain through. It was necessary to interrupt the observations, and in January 1949, to dismantle the refractor. André Danjon (the director of Paris observatory) then offered me to go to the east tower, and to use the 38 cm equatorial."

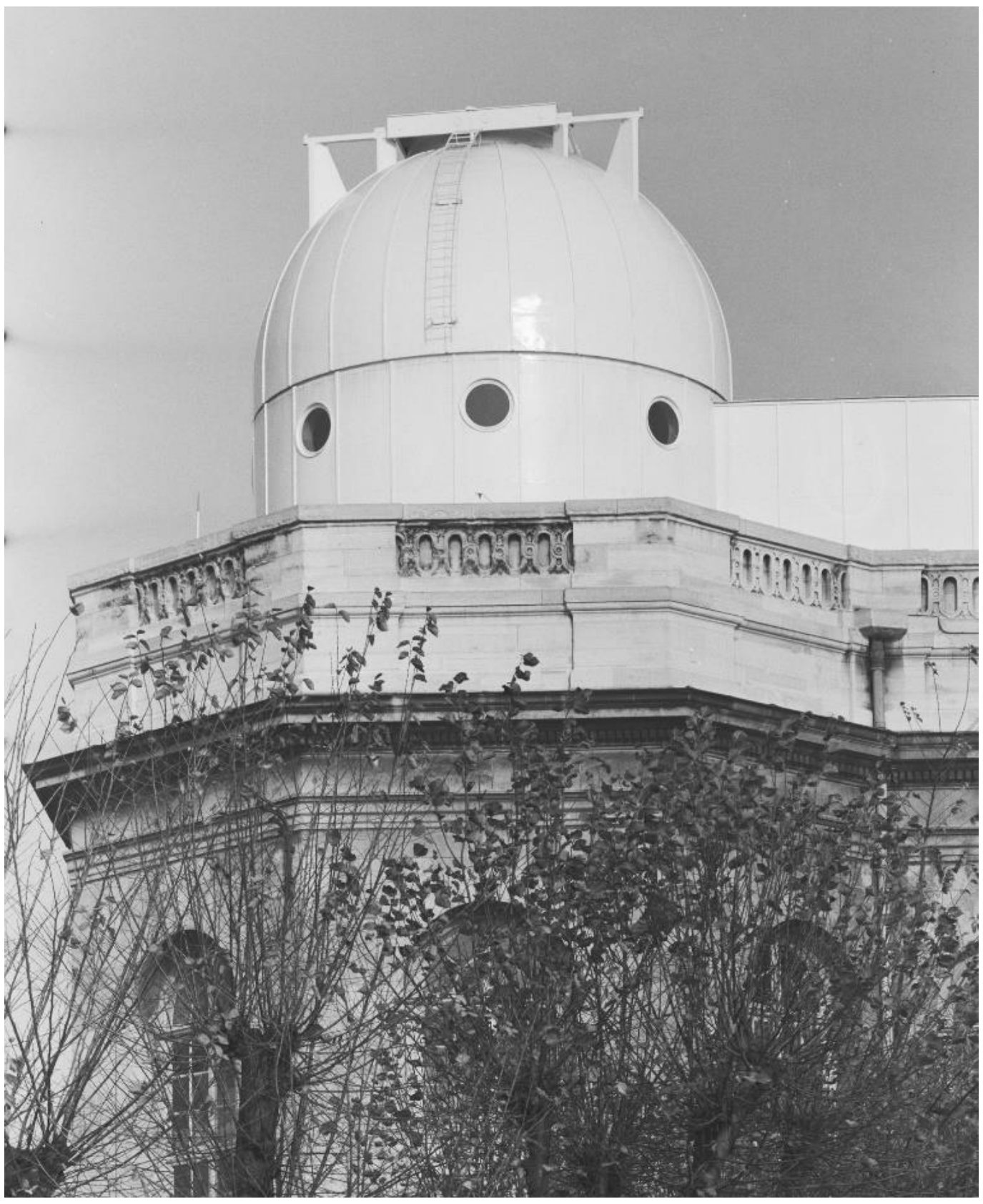

*Figure 2: the new metal dome of the west tower (courtesy Paris observatory)*

The wooden dome of 1858 was abandoned and replaced by a new metal dome (figures 2 & 3) in the course of the fifties; this is the one transferred to Saint Véran in 1974. We have not been able to identify its manufacturer in the archives; it cannot be Eiffel, as has sometimes been read, but the hypothesis of one of the companies that succeeded the Eiffel workshops cannot be ruled out. An equatorial table was placed there, as shown in figure 4.

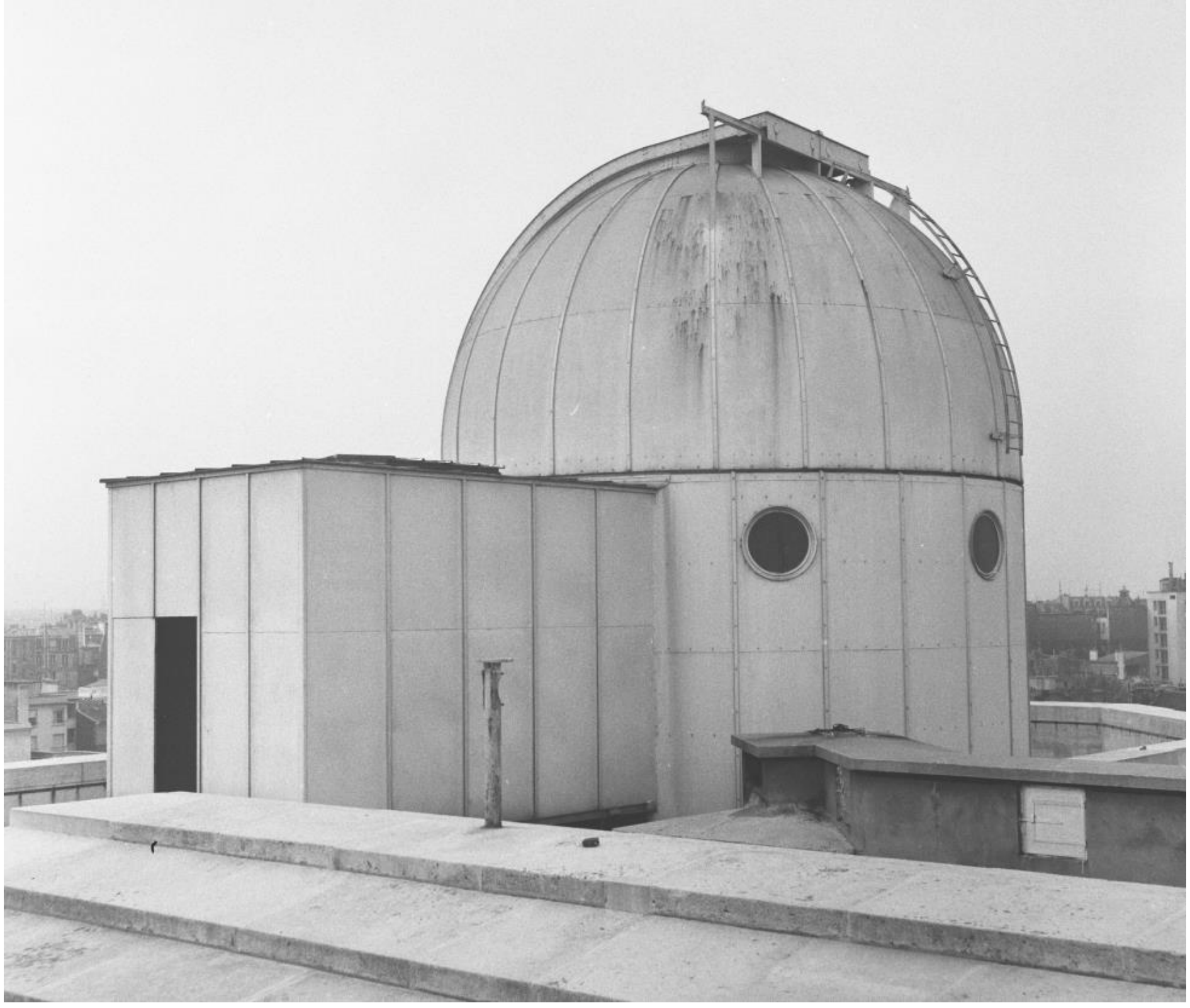

***Figure 3****: the new 7-metre metal dome of the west tower in September 1961 (courtesy Paris observatory)*

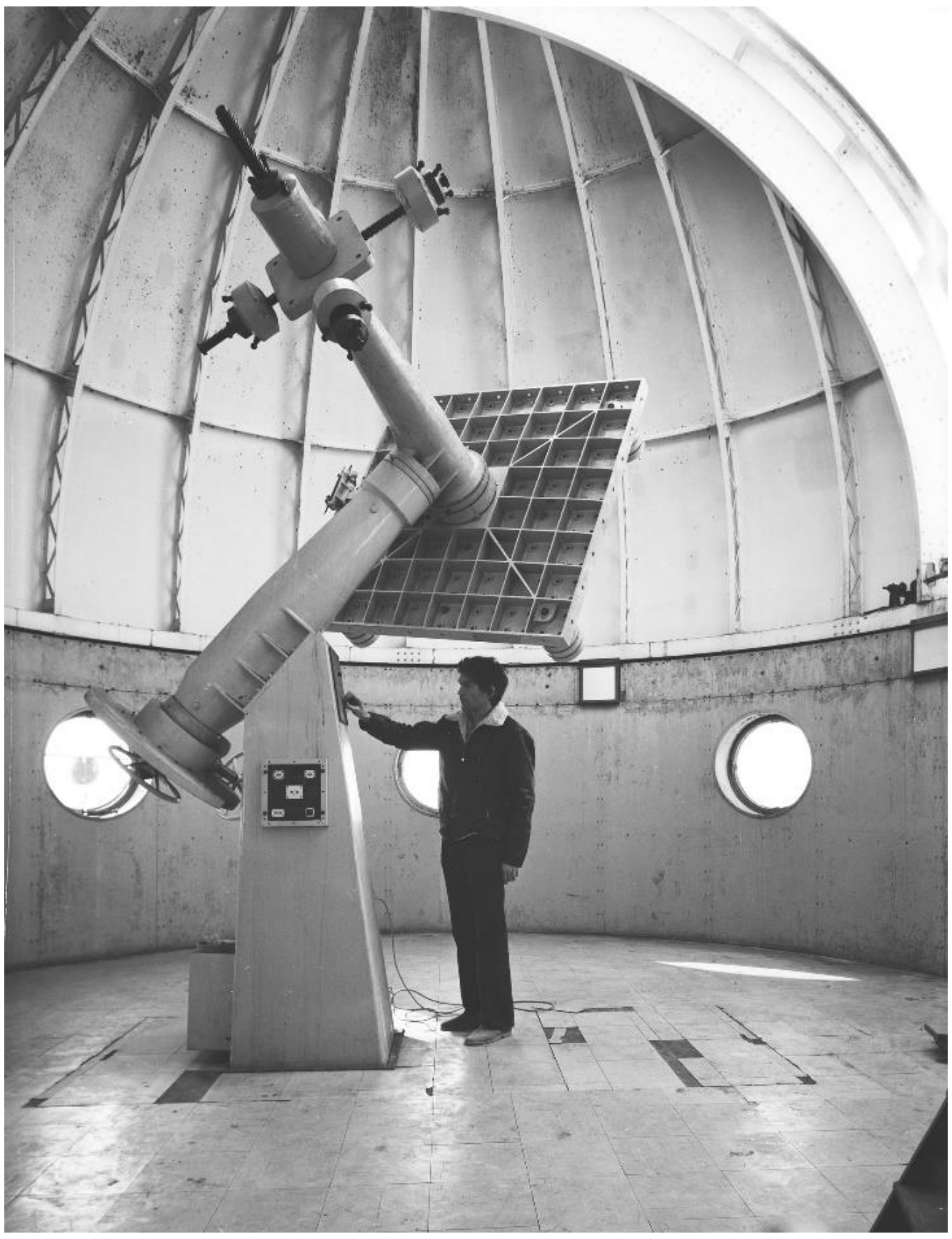

***Figure 4****: the equatorial table of the west tower in May 1974 (courtesy Paris observatory)*

## THE TRANSFER OF THE DOME FROM THE WEST TOWER TO SAINT VÉRAN IN 1974

The selected place was "Pic de Château Renard" (2930 m) above Saint Véran village. The "Hautes Alpes" administrative district took charge of the construction of the access track in June 1974 (only usable in summer by 4-wheel drive). The Paris Observatory provided the dome of the west tower, and the INAG paid the sheet metal housing hut and the generators (because the connection to the public electric network was impossible). The Nice Observatory lent an equatorial table for the dome. The solar instrument (a coronagraph of 25 cm and 3 m focal length) was installed in October 1975. The 7 m diameter and 9 m high dome of the Paris observatory required the construction of a masonry foundation. Under the observation floor, there was a basement used as a workshop. A 5 m high metal annex, also with two levels, was attached to the dome. The opening of the cupola was 2.30 m wide and it rotated over 360°, both motions were manual. The dismantling, transport, and reassembly took two years (1974-1975). Gabriel Rousset (Meudon workshop) wrote in 1981: "We were in the presence of a rusty dome, having suffered from the dismantling and reassembly spread over two years, with a rotating hard spot and a slight misalignment of the closing of the trapdoors; and zero impermeability to snow and condensation. To eliminate these defects, it was necessary to do many tasks, such as redo the painting, realign the rotation rollers to reduce the hard spot, eliminate snow infiltration with a system of joints and skirts, deposit a waterproof coating on all joints and spray polyurethane on the inner wall of the dome to prevent water condensation. This work was spread over two summer campaigns, by which we had a more or less suitable dome, always having a defect in rotation and in closing the trapdoors".

In September 1977, the station (figures 5 to 15) included:

- The dome and its annex, plus the coronagraph (described in the next section).
- A metal sheet house of 40 m² used as a living base, including kitchen, living room, bedroom, bathroom, which could accommodate 4 people. It was heated by gas (12 bottles of 37 kg); water (non-drinkable) was delivered by tanks (14000 litres), and electricity by burning fuel generators.
- a second metal hut of 15 m² used as a warehouse.
- a technical hut containing 3 generators of 4, 6 and 10 KVA, as well as a small group of 1 KVA for domestic use; the fuel tank (3000 litres) was installed underneath.
- An iron tunnel connecting the dome to the living base and the warehouse.
- a telephone line and a safety radio link.

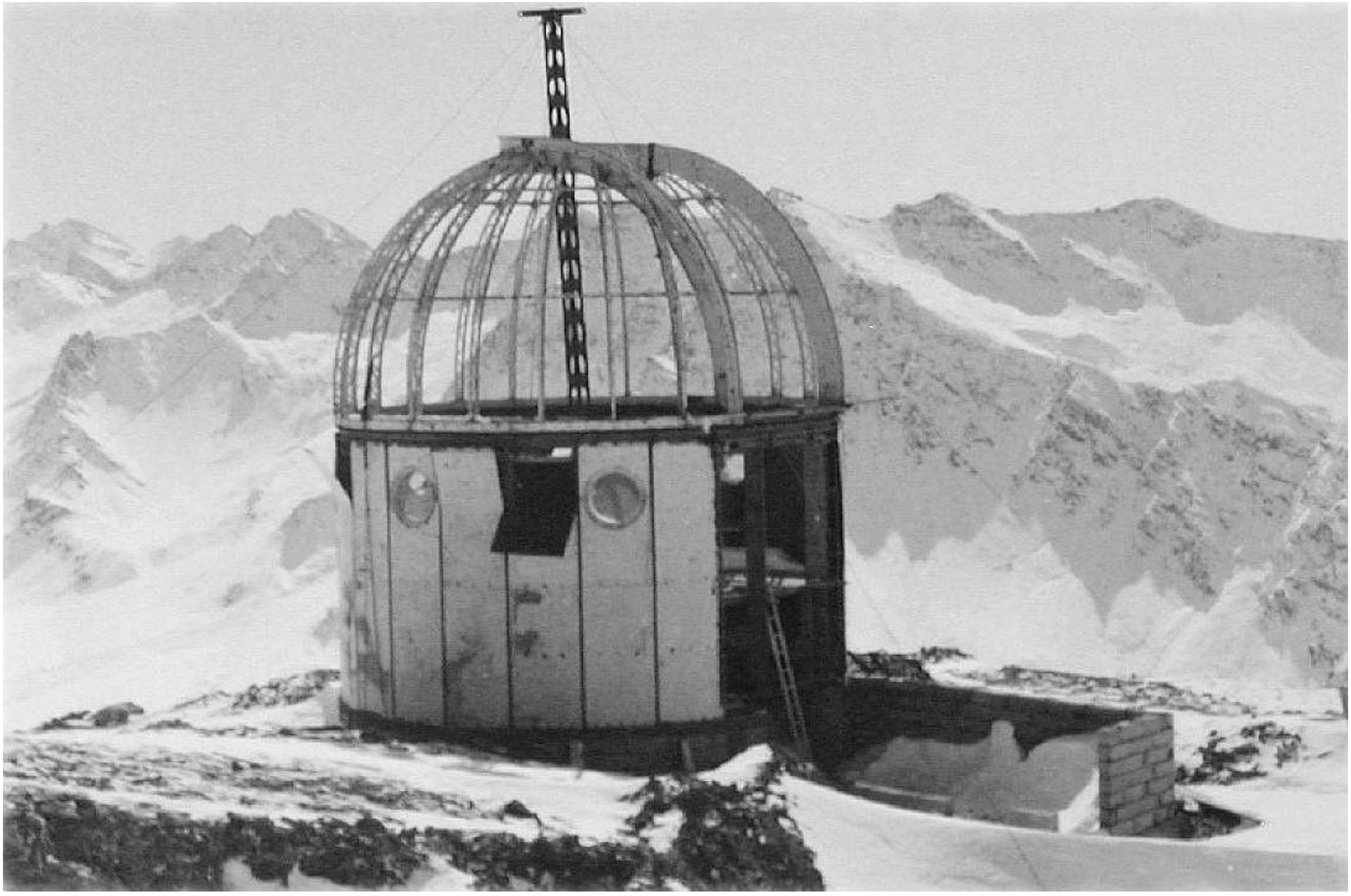

*Figure 5: the reassembly of the 7-m dome during 1974 and 1975 (courtesy Paris observatory)*

The assembly of the station, dome and coronagraph required the active participation of many people, who are quoted by Bellenger in his web site: Rémy Bellenger (optics, electronics), Yves Zéau (logistics), Jean-Pierre Dupin (electronics), Gabriel Rousset and Isidore Raulet (mechanics), Annick Fayet, Jean Guérin and Jacques Baudrand (instrumentation), Mireille Dentel (computing), Christiane Jouan (administration). Many people of Saint Véran village (Joseph Brunet, Jacques Jouve, Pierre Prieur-Blanc) helped a lot, and Jean-Eugène Chabaudie from Nice observatory. Guérin was responsible of the instrument from 1975 to 1977 and Bellenger from 1978 to 1982.

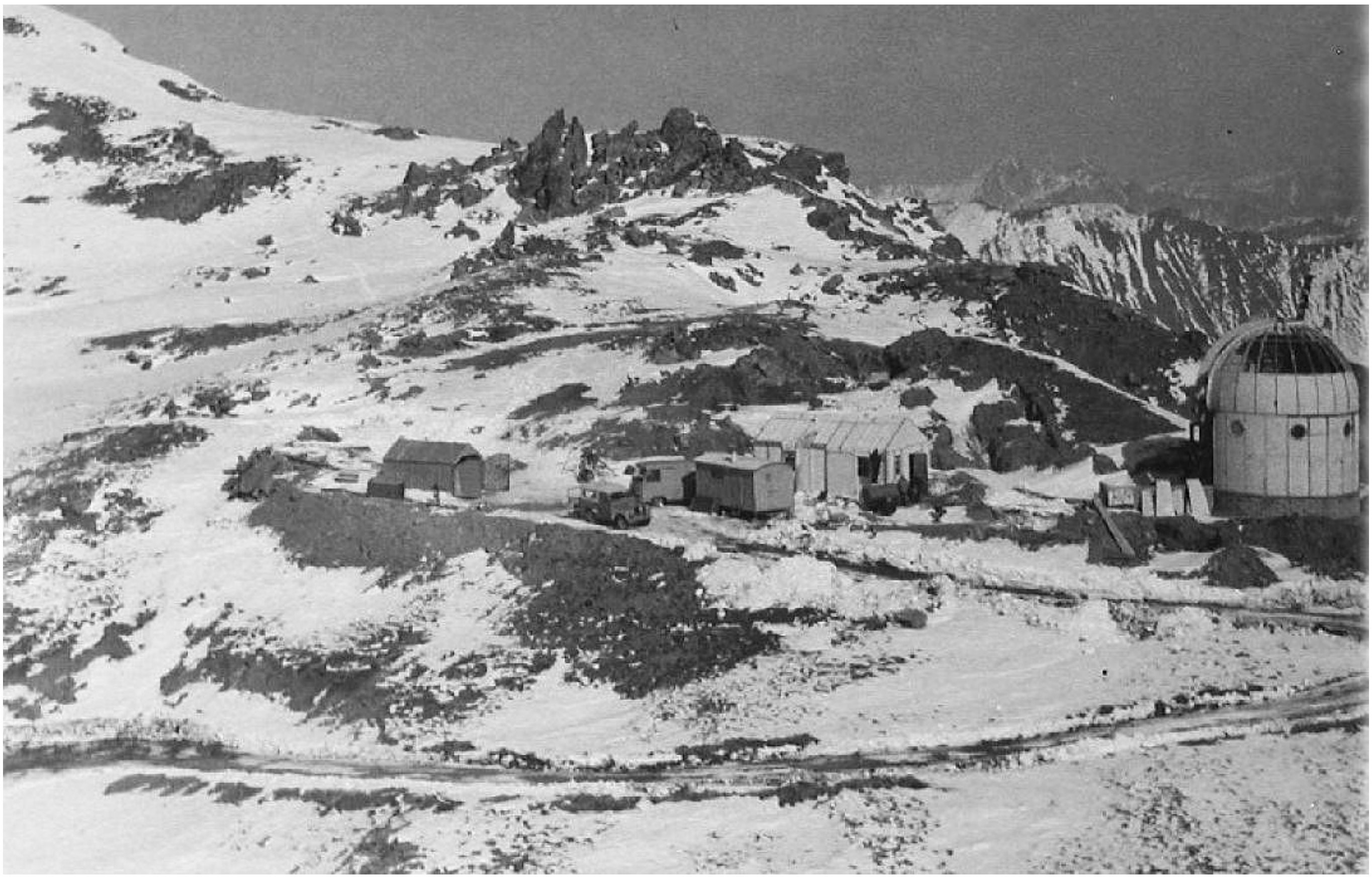

*Figure 6: the station under construction during 1974-1975 (courtesy Paris observatory)*

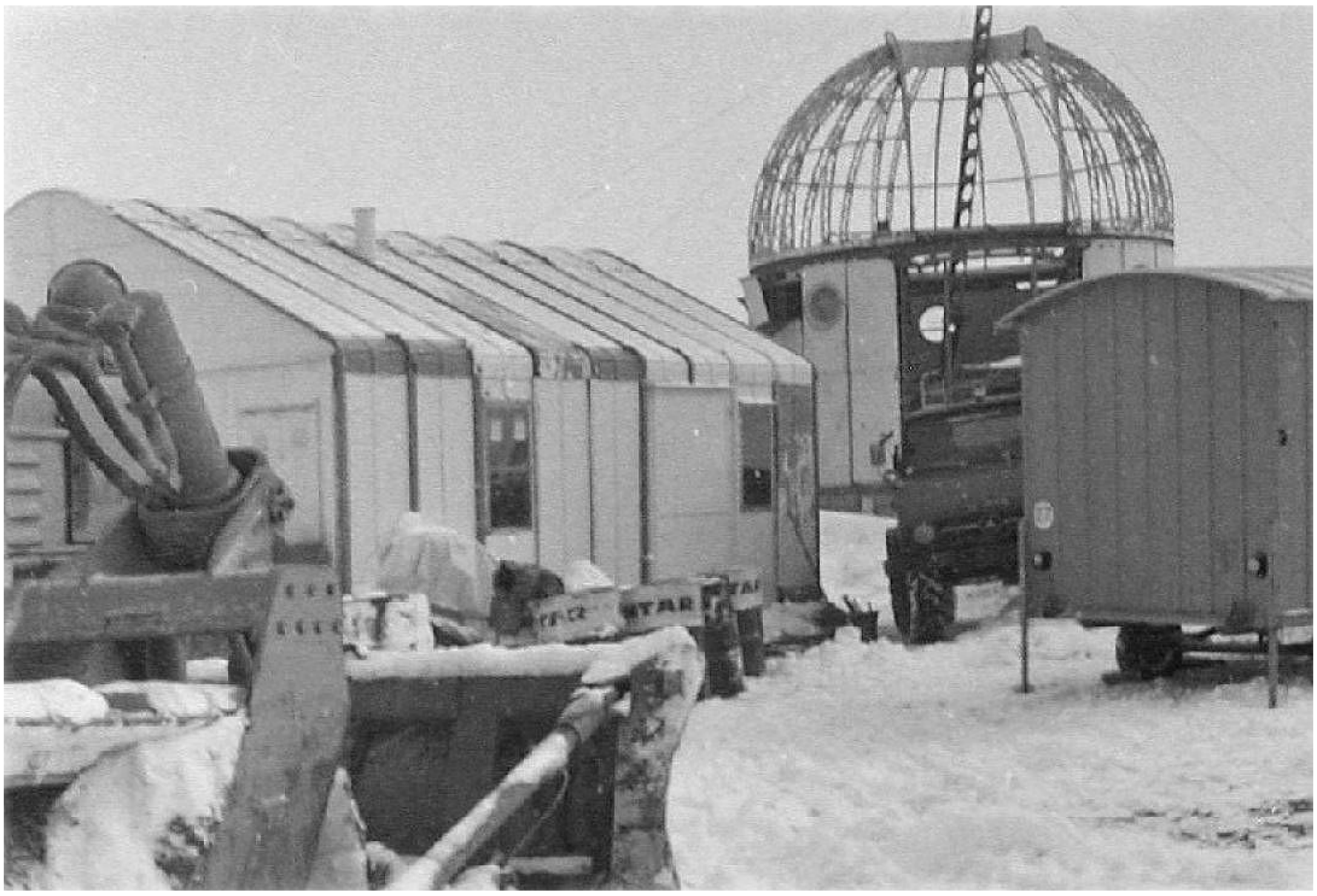

*Figure 7: the living area (left) and dome (right) under construction in 1974-1975 (courtesy Paris observatory)*

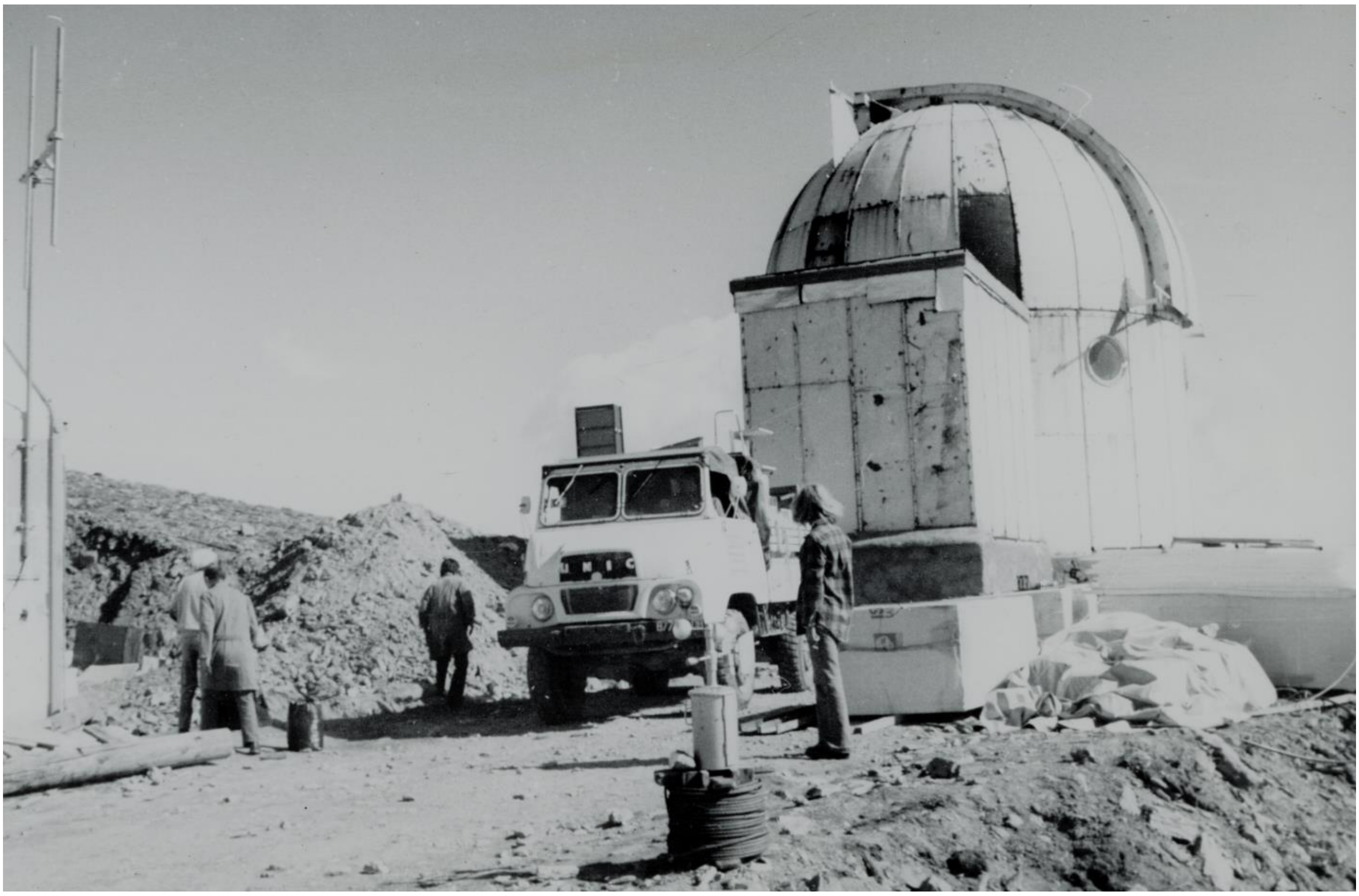

**Figure 8**: preparing the tunnel between the house (at left) and the dome in 1975. The UNIC 4 x 4 truck was the one already used for the eclipse mission of 1973 in Mauritania (courtesy Paris observatory)

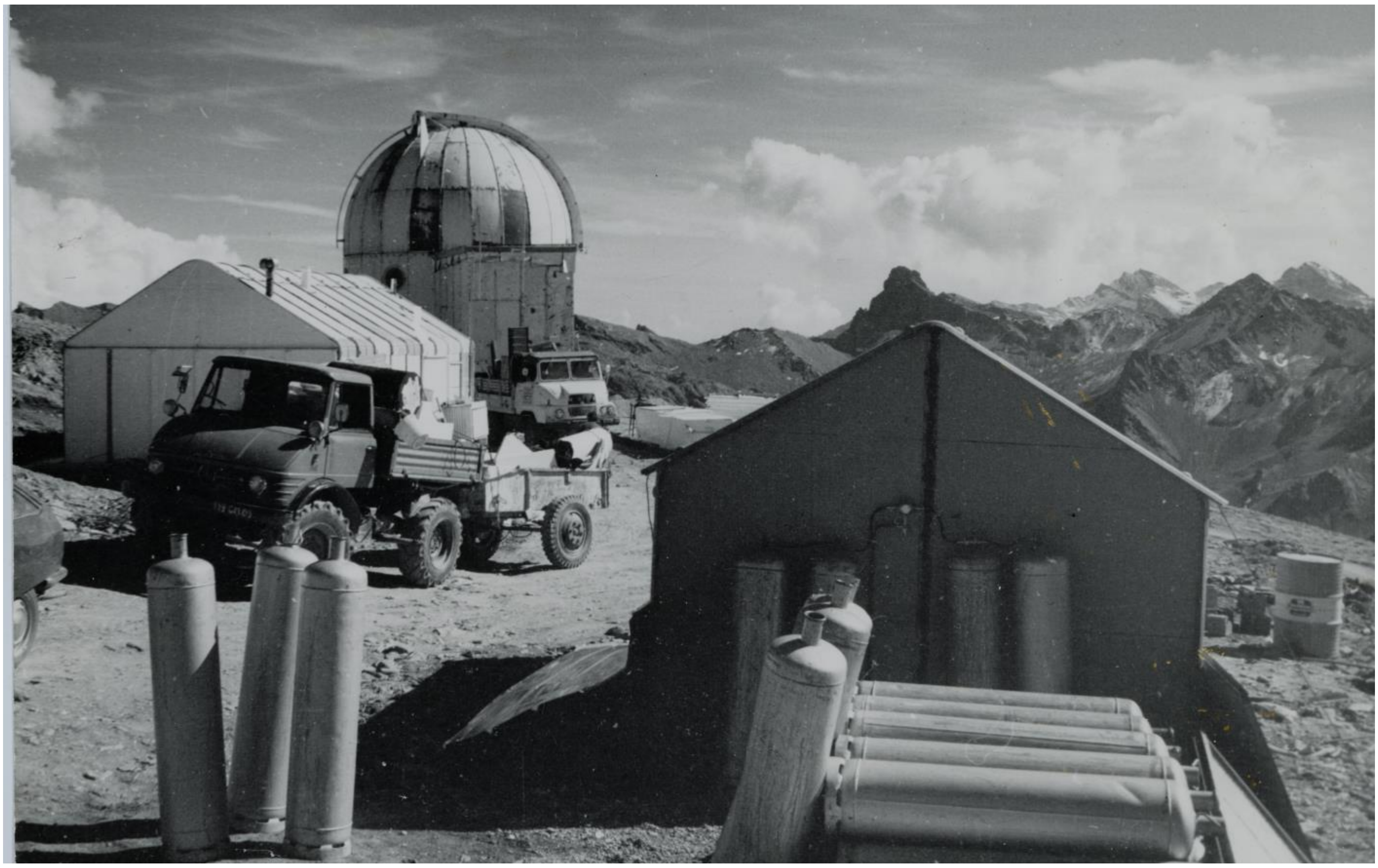

**Figure 9**: building the station in 1975; the heating of the house was provided by gas and electricity by fuel generators. The Mercedes Unimog 4 x 4 truck belongs to Joseph Brunet, Saint Véran (courtesy Paris observatory)

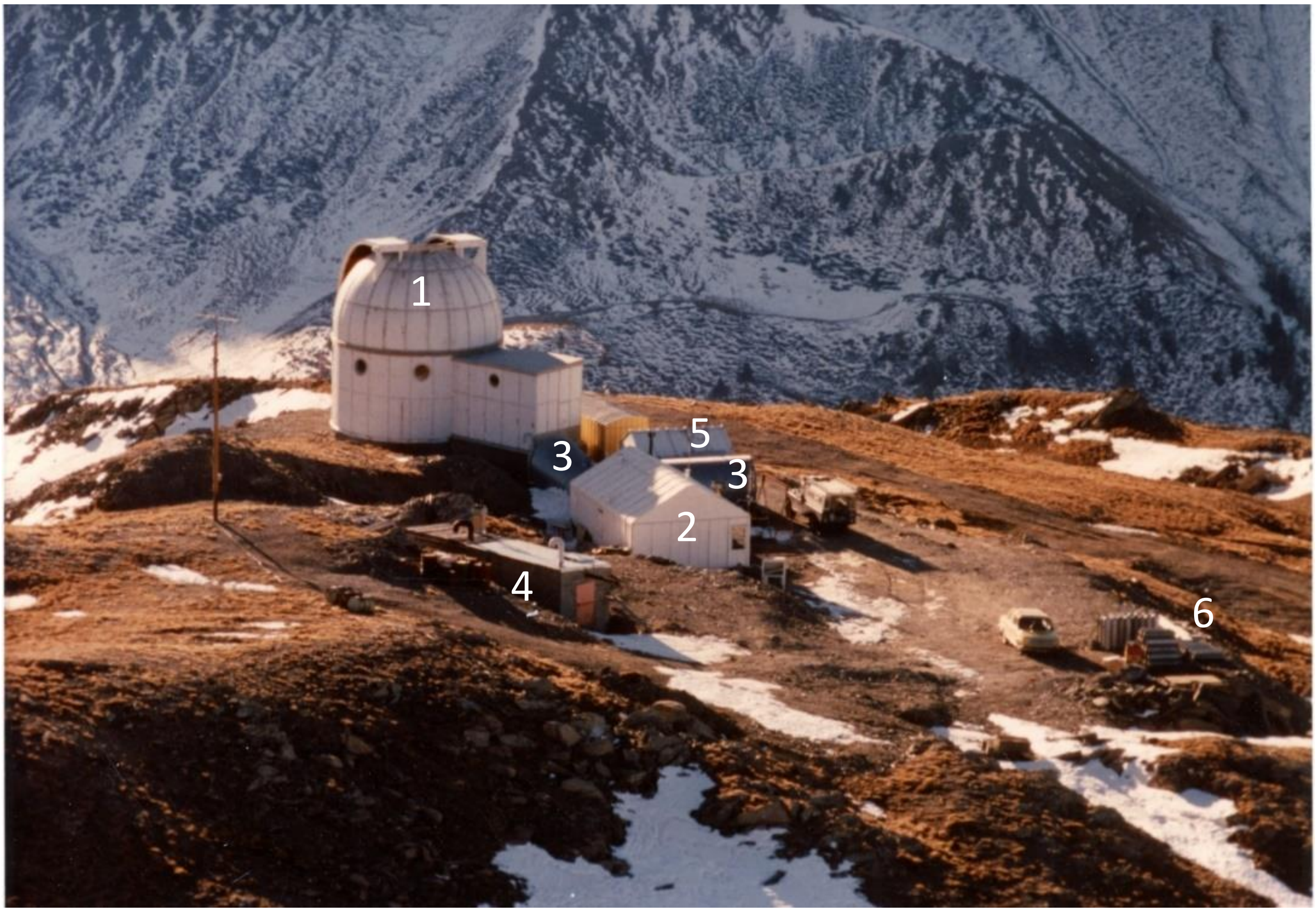

*Figure 10: the station in 1977; one sees the 7-metre dome and two layer annex (1), the house (2), the double tunnel (3), the electric plant (4), the warehouse (5) and gas storage area (6) (courtesy Paris observatory)*

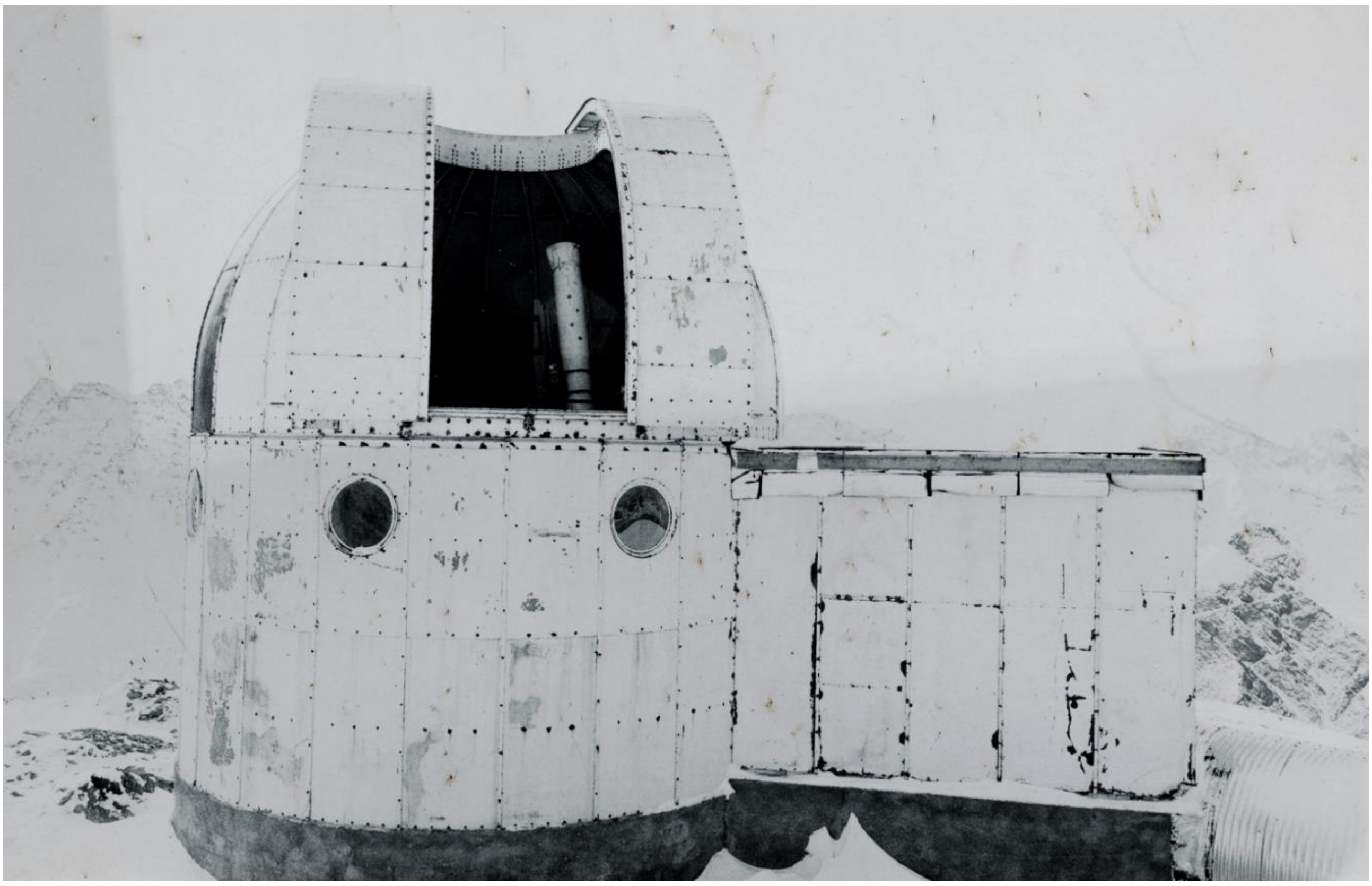

*Figure 11: the dome and annex, end 1975, before being repainted (courtesy Paris observatory)*

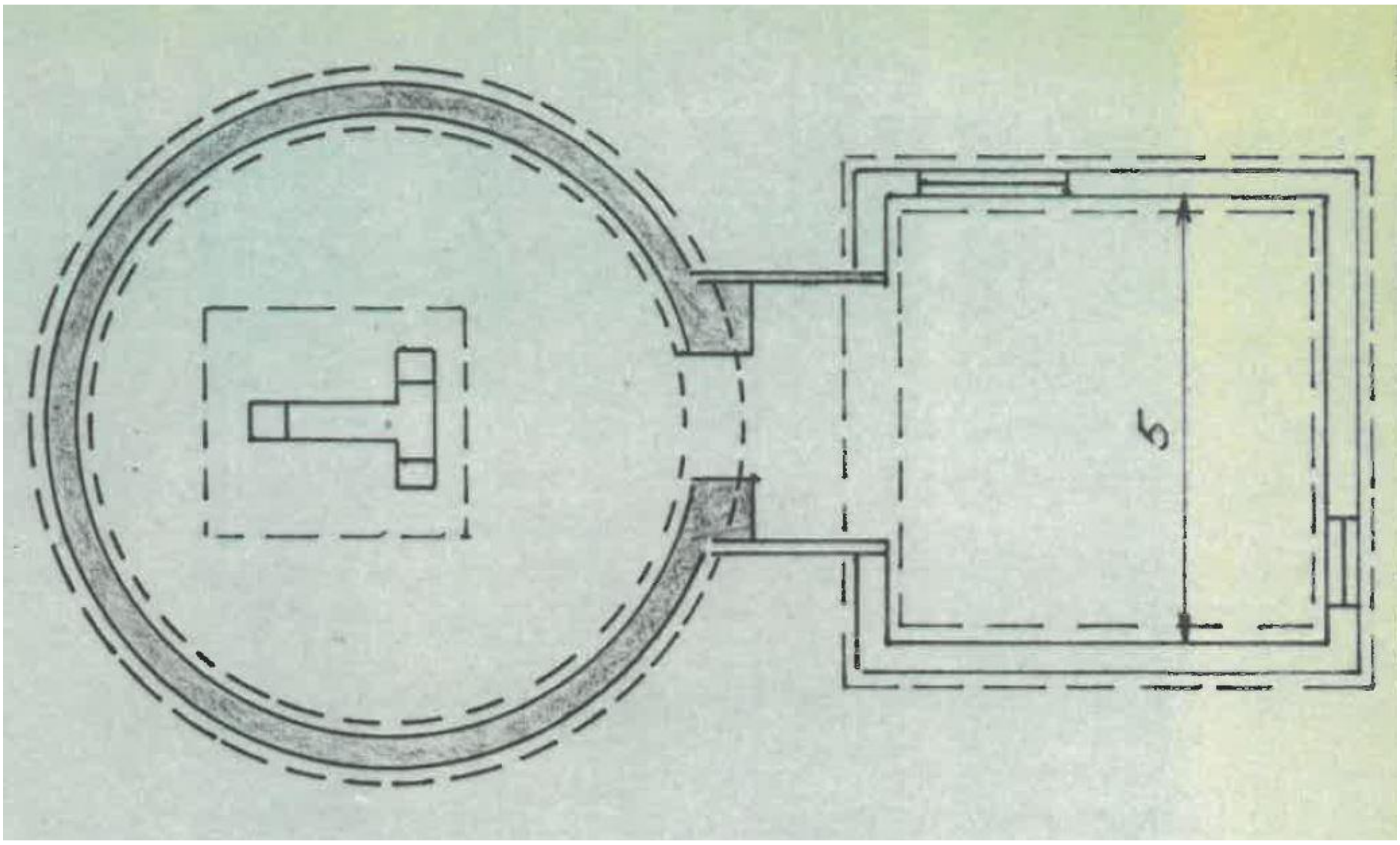

*Figure 12: plan of the dome and its annex from above (courtesy Paris observatory)*

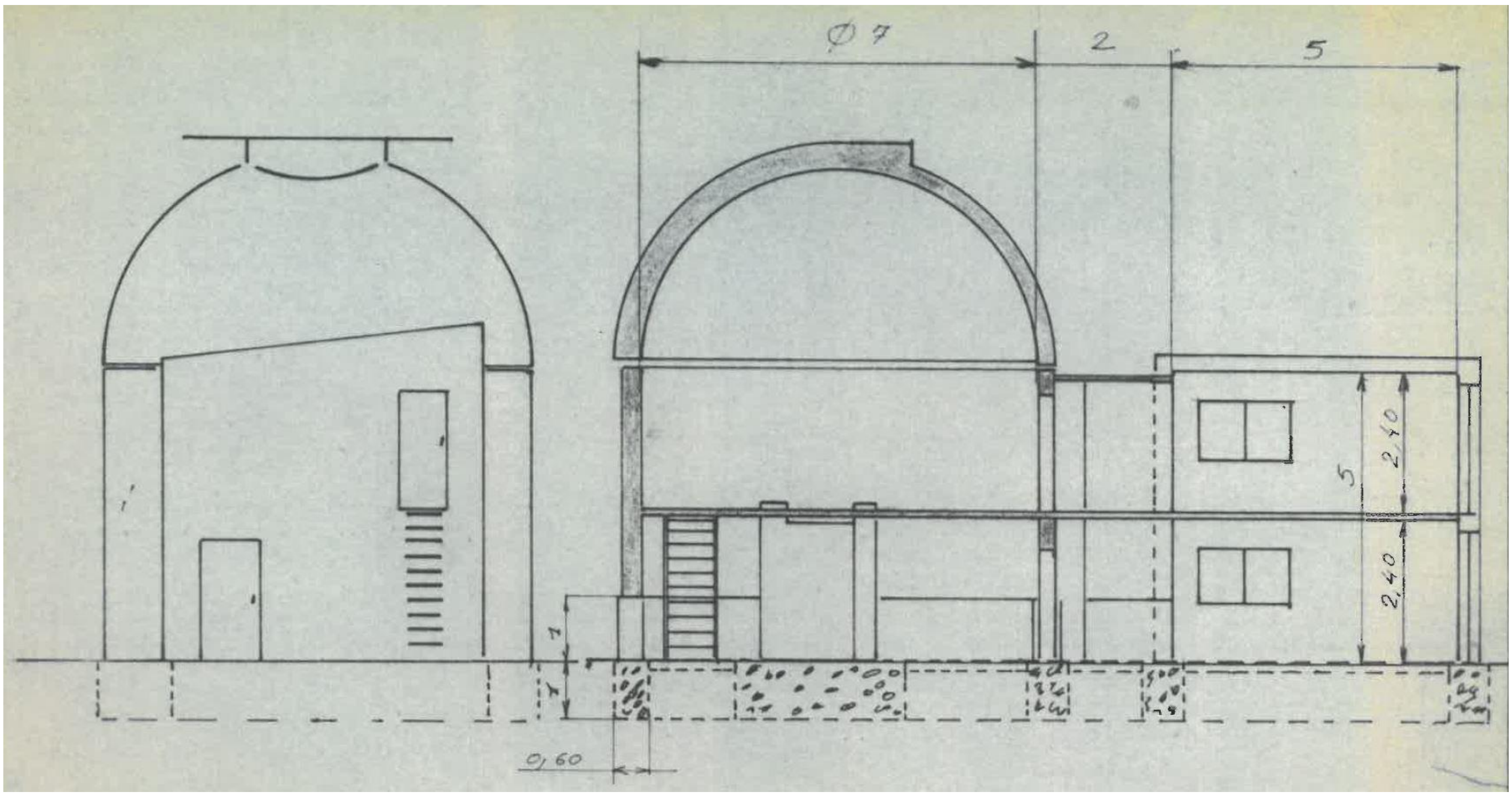

*Figure 13: side view (left) and cross section (right) of the dome and its annex, with two levels including a workshop below the observing floor. The height of the dome was 9 m, and its diameter 7 m. The technical annex was a square of 25 m², two floors, 5 m high (courtesy Paris observatory)*

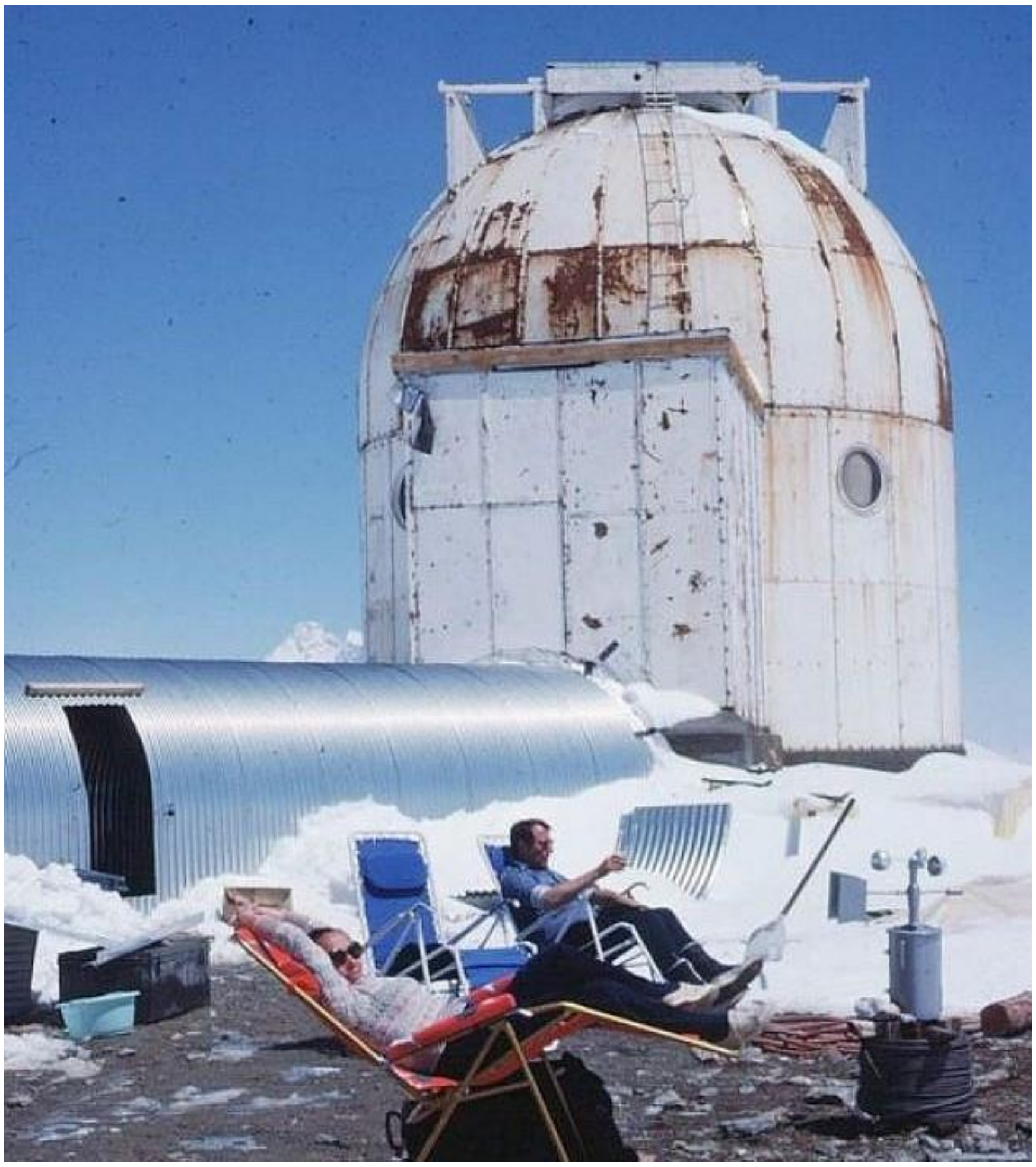

*Figure 14*: *the dome, technical annex and tunnel towards the living area in 1976 (courtesy Paris observatory)*

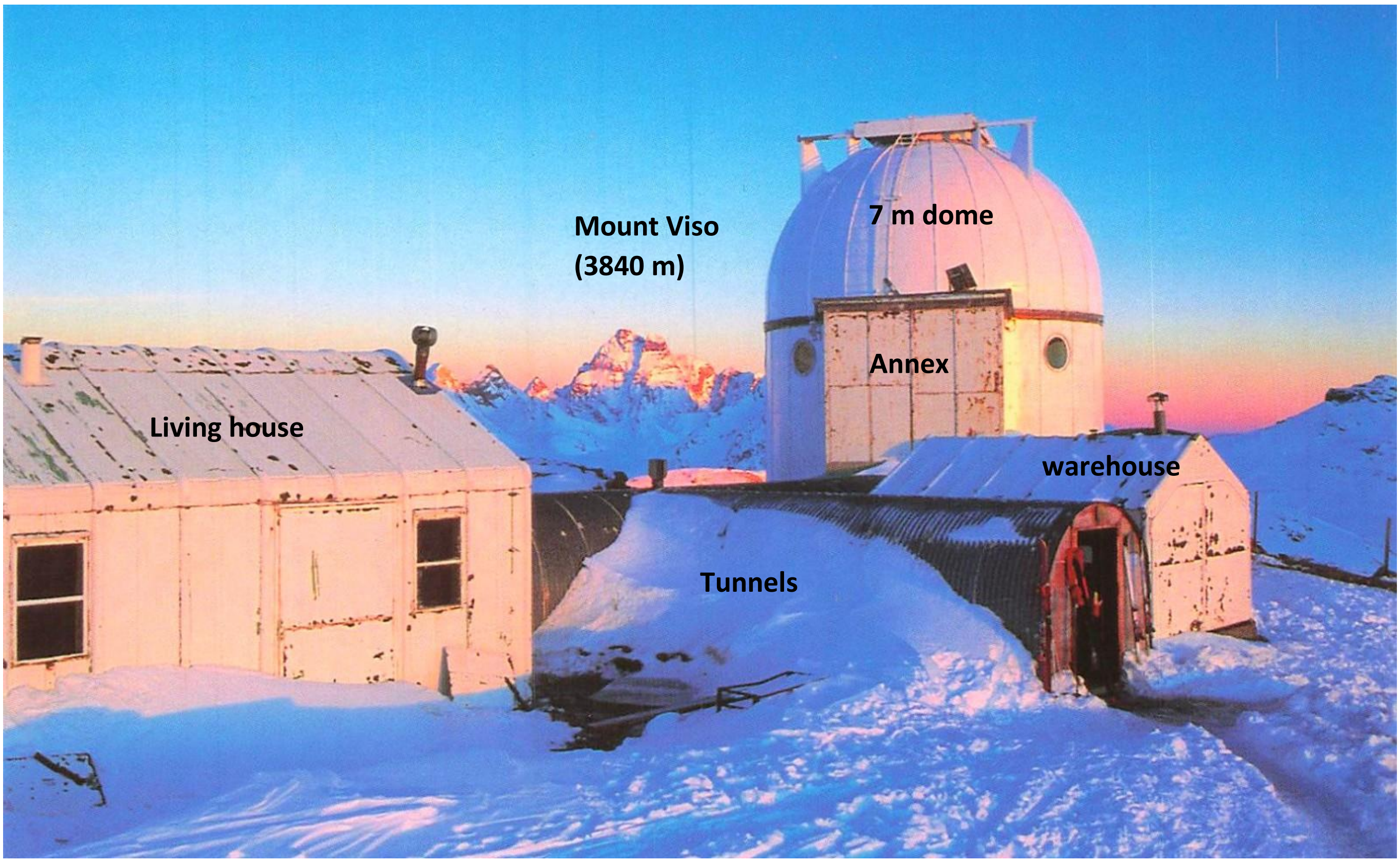

*Figure 15*: *a postcard of the station (2930 m) at the end of the seventies*

## THE LALLEMAND ELECTRONIC CAMERA AND SOLAR OBSERVATIONS

Despite of the astronomical quality of the Pic de Château Renard at Saint Véran, identified as the best nocturnal site of France by INAG in 1970-1971, the site of Hawaii (Mauna Kea at 4200 m) was selected in 1972 for the construction of the Franco-Canadian telescope of 3.60 m diameter (the CFHT). At the same time, when CCD targets did not exist, coronagraphy projects, based on the Lallemand electronic camera, were in progress: its principle (Lallemand, 1937; Lallemand *et al*, 1960) is based on photoelectricity (photon/electron conversion working with a photocathode), then electrons are focused by electronic lenses on a nuclear photographic plate, which made the Lallemand camera 100 times more sensitive than conventional photography (close to modern CCD). It was therefore well suited to the observation of faint objects (such as galaxies). The first assembly of an electronic camera on a coronagraph associated with a spectrograph (Figure 16) was made by Rozelot & Despiau (1972) at the Coupole Tourelle, Pic du Midi (Raymond Despiau was also a famous Pyrenean climber, who took part to the French expedition at Everest, headed by Pierre Mazeaud in 1978). The first observations of a total eclipse with an electronic camera were conducted by Fort *et al* (1972) in Mexico (1970). Two years later, the Meudon camera group lined up two electronic cameras at the 1973 eclipse in Mauritania (Fort & Picat, 1975, Picat *et al*, 1979), after having carried out tests at the top of the Auron cable car (Southern Alps, figure 17); one with electrostatic focusing for imaging in forbidden coronal lines, conducted by Bernard Fort, and the other one with magnetic focusing, conducted by Jean-Pierre Picat, for spectroscopic studies (Figure 17). It appeared natural to the Meudon group, led by Paul Felenbok, to continue this strong technical and scientific investment with a high altitude coronagraph at Saint Véran, a good site in perspective, but where the station had to be created from nothing, an ambitious challenge !

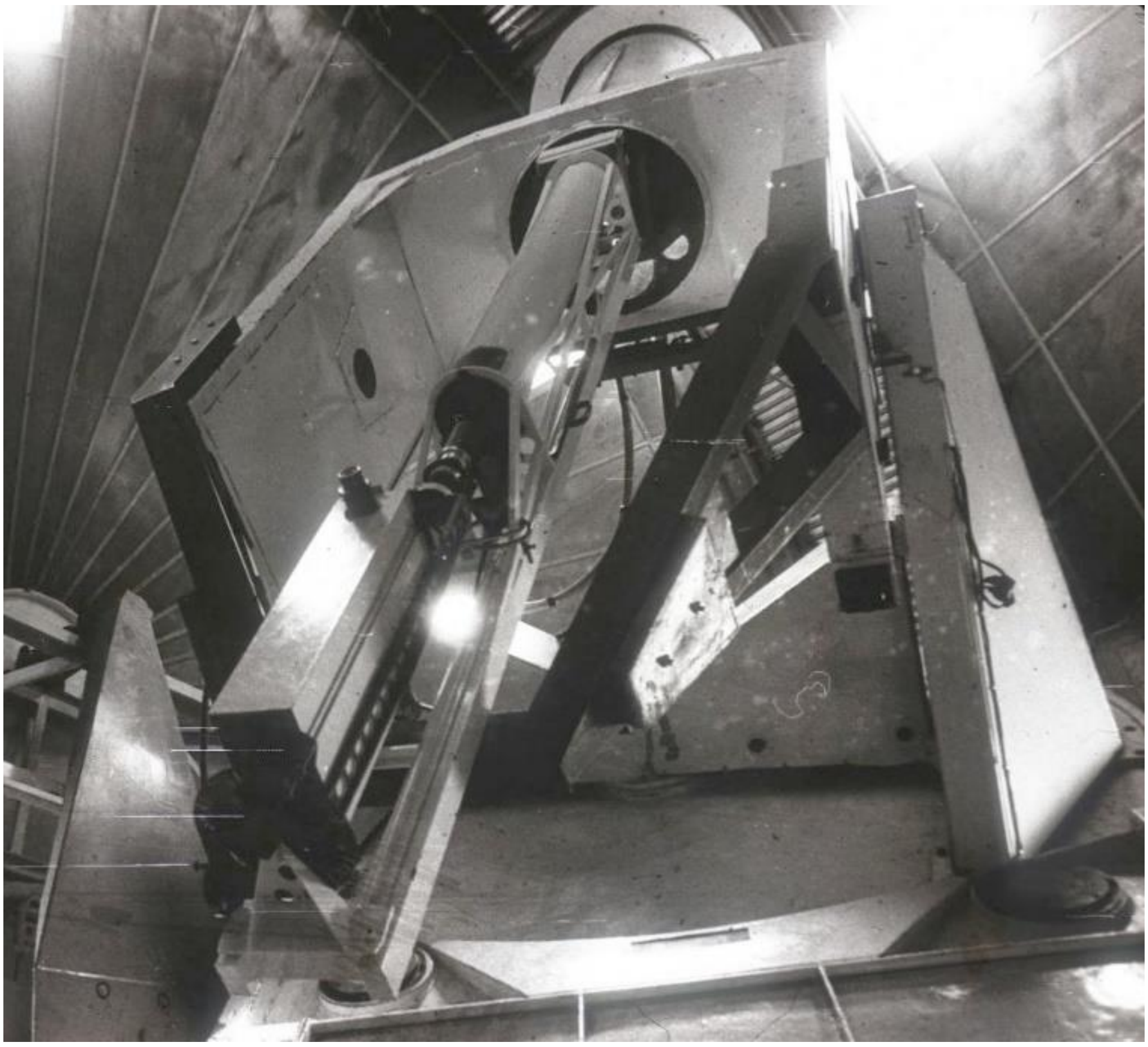

***Figure 16**: the 15 cm spectro-coronagraph of Rozelot (1972) was the first to work (occasionally) with a Lallemand electronic camera, at the Pic du Midi (camera not mounted on the photo, courtesy Rozelot)*

It should be noticed that Felenbok's team had developed a version of the Lallemand system, the electronic valve camera (Baudrand *et al*, 1972), which avoided breaking the vacuum during the change of nuclear plate. Indeed, the issues of vacuum and cooling by cryostat at low temperature (liquid Nitrogen, 77 K) were major obstacles to the use of these delicate devices, especially in environments with reduced logistics (such as Saint Véran or eclipse missions). As solar structures may evolve rapidly (the minute for coronal mass ejections), it was essential to be able to change plates quickly.

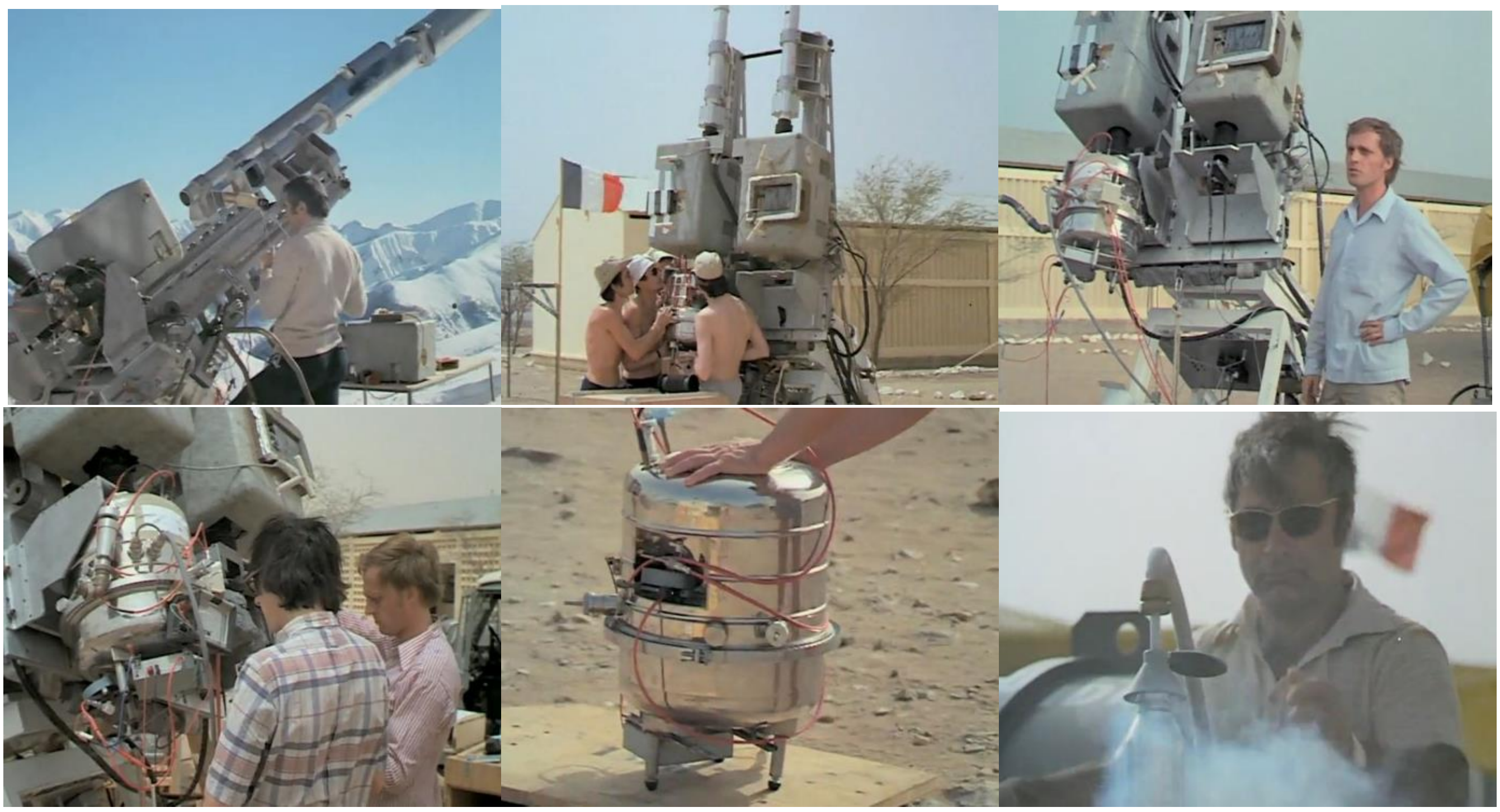

*Figure 17: the electronic cameras of Felenbok's group tested in Auron (French Alps, top left) and then used in Mauritania for the 30 June 1973 eclipse. Bottom right: Paul Felenbok in Atar (courtesy CNRS media)*

**THE PRINCIPLE OF THE CORONAGRAPH**

It was invented by Lyot (1932). The principle was described more simply by Crifo (1981). It is a much more subtle instrument than a classical refractor with a simple occulting mask in the image plane to simulate an eclipse. Indeed, the corona is $10^6$ times less luminous than the solar disk, whose light penetrates the coronagraph (this is not the case during an eclipse) and produces parasitic light. First of all, there is the diffusion by the dust of the sky (attenuated in the high mountains) and the molecules, as bright as the corona; it will therefore be better to observe towards the red with filters, because Rayleigh scattering, in $1/\lambda^4$ (where λ is the wavelength), dominates in the blue. Then we are confronted with the reflection of light on the inner faces of the entrance lens (Figure 18), which is also brighter than the corona. At the primary focus, in the image plane, a reflecting cone obscures the solar disk; but there is a secondary focus, resulting from two internal reflections (attenuation $R = 0.04^2 = 0.0016$). The secondary focus has a very short focal length, equal to $[(n-1)/(3n-1)]\ F = F/7$ for a refractive index $n = 1.5$, if F is the focal length of the primary focus. At last, the diffraction of light by the contours of the objective, which is also brighter than the corona, must be eliminated in the pupil plane.

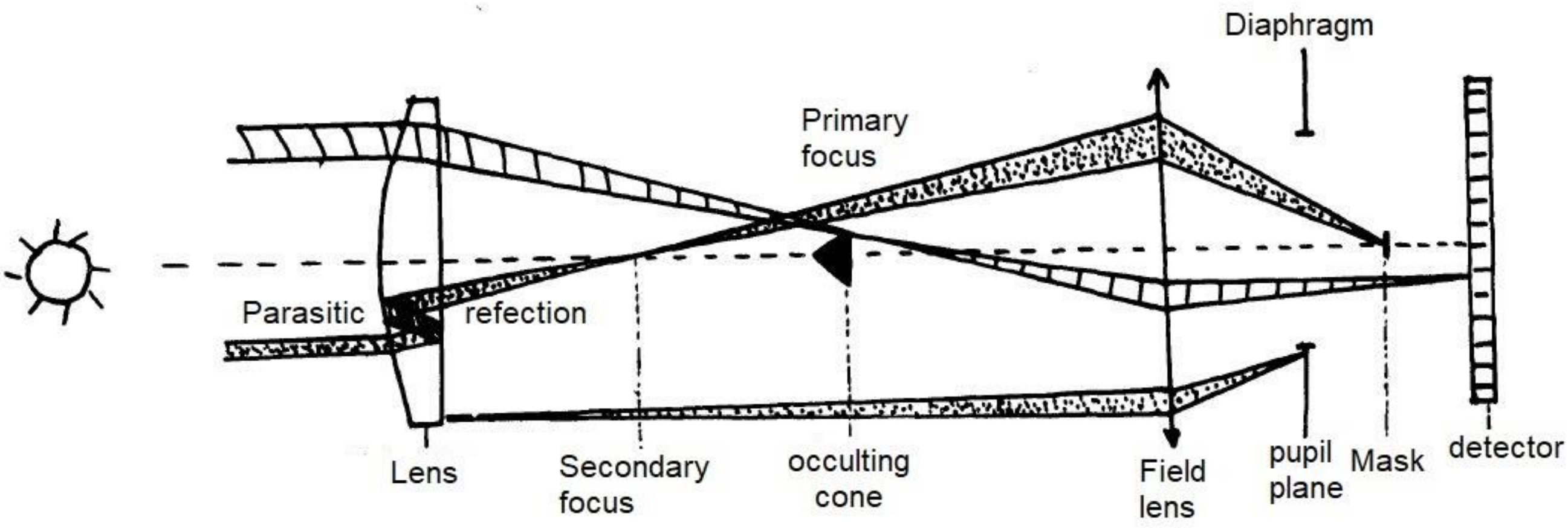

*Figure 18: diagram of a coronagraph (after Crifo, 1981). The occulting cone is at the primary focus. The secondary focus forms by two internal reflections. The field lens gives a pupil image which is cut by a diaphragm to eliminate edge diffraction. There is a small circular mask (the Lyot stop) to cut the secondary image.*

How to eliminate these parasitic sources ? Near the occulting cone, in the vicinity of the primary focus, a field lens is placed, which forms an image of the entrance objective in the pupil plane. At this location, a diaphragm cuts the diffractive edges of the pupil (figure 19). Just behind this diaphragm, there is the image of the secondary focus by the field lens, it is a small central luminous spot of a few millimetres (figure 19), which can be masked by the Lyot stop. After eliminating these major defects, an image of the corona forms on the detector; a photographic plate for the low corona is sufficient, but an electronic camera (or a CCD today) is necessary for the far corona.

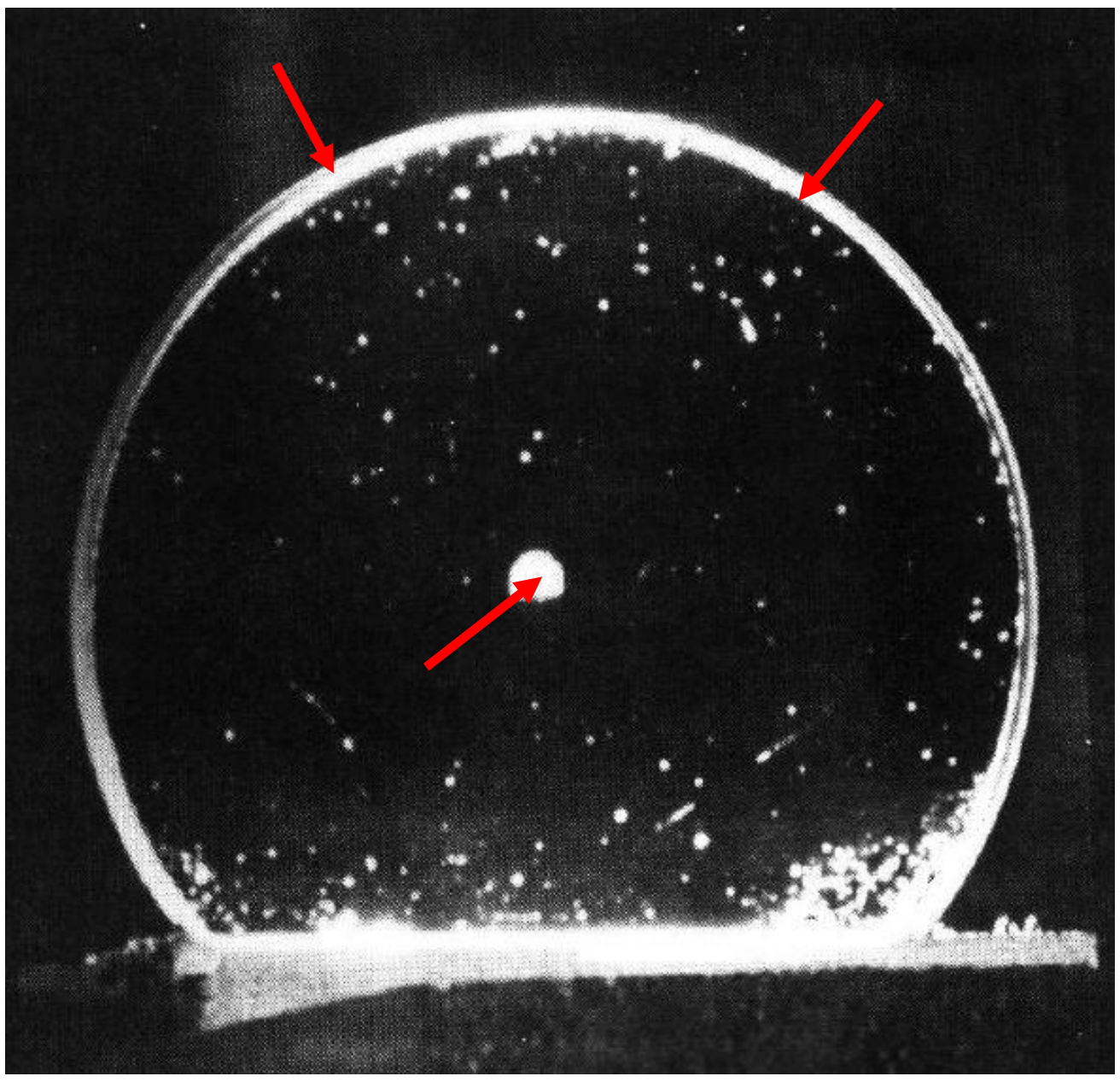

***Figure 19****: Image of a good quality coronagraph objective (After Dollfus, 1983). We can distinguish the diffractive edge and the central reflection (both are 1000 times brighter than the corona, they are eliminated by placing a diaphragm in the image of the pupil, and by obscuring the image of the secondary focus by the Lyot stop). There are also dusts, heterogeneities, scratches and bubbles, which scatter light in the lens glass. In order to minimize the defaults, a single lens of exceptional quality and precise polishing is necessary. This implies observations with narrow filters because of the chromatism of the glass, which cannot be corrected. A mirror telescope would be much more diffusing than a single lens refractor.*

**THE CORONAGRAPH OF SAINT VERAN**

Paul Felenbok left us an optical diagram (Figure 20). The objective is a single lens with a diameter of 25 cm and a focal length of F = 300 cm (Figure 21). The solar image at the primary focus has a diameter of 2.8 cm and is intercepted by the reflecting cone. At this place (the image plane), there is a field lens with a focal length of 85 cm. It forms an image of the entrance pupil about 120 cm away, where the diaphragm and the Lyot stop are located. The diaphragm reduces the coronagraph diameter from 25 cm to 16 cm by eliminating the diffractive edge. The secondary image focus (F/7 = 43 cm), resulting from the internal reflections in the entrance lens, is 43 cm away from it and gives a small (but bright) parasitic solar image of 0.4 cm diameter; its image through the field lens, just behind the diaphragm, is even smaller and obscured by the Lyot stop.

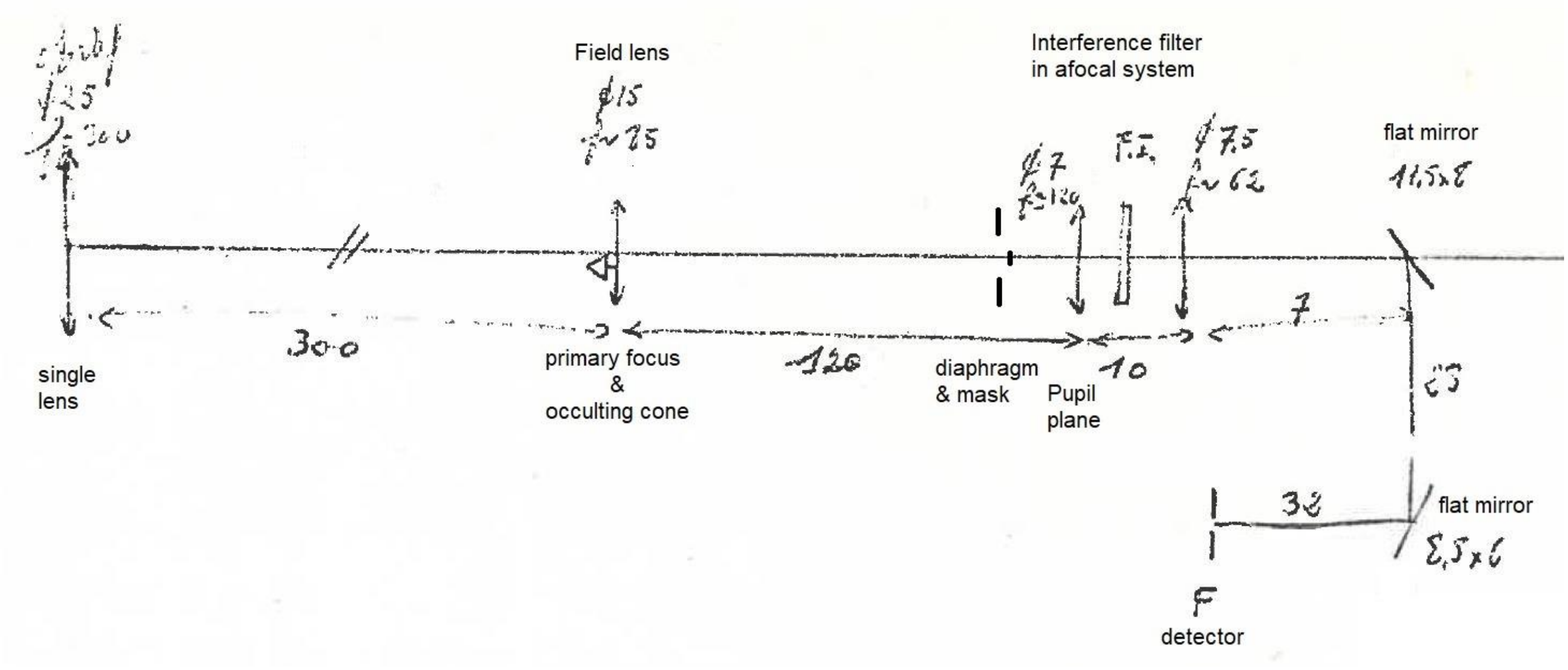

***Figure 20****: diagram of the coronagraph of Saint Véran (after Paul Felenbok’s original drawing).*

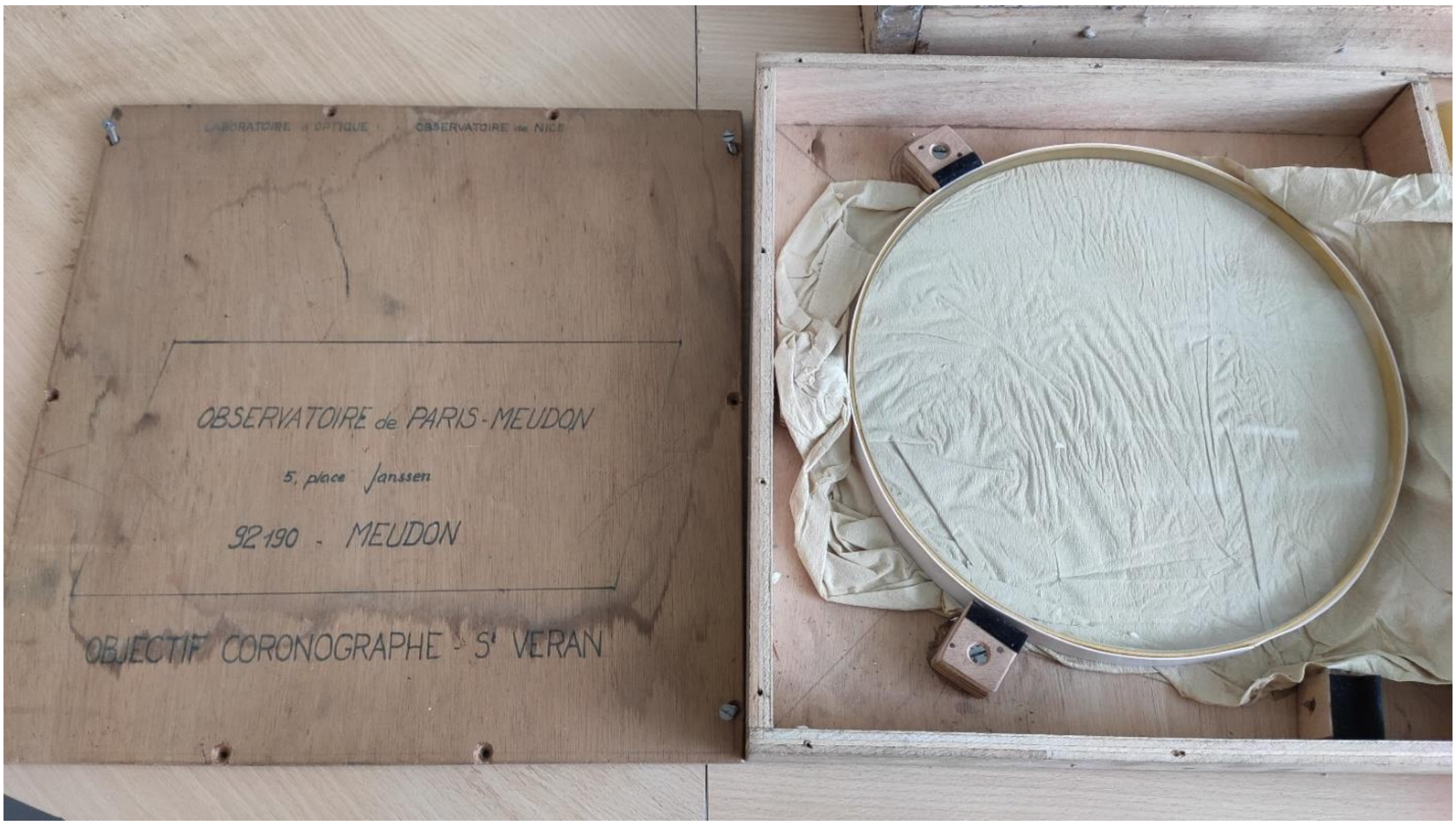

***Figure 21**: the coronagraph lens with a diameter of 25 cm and a focal length of 300 cm, polished by Demarcq (former optics laboratory of Nice Observatory). It is a convex (R = 1.5 m) planar single lens (courtesy Paris observatory).*

This is followed by a train of two lenses containing an interference filter isolating a coronal emission line, a few Angström wide. This afocal system consists of two converging lenses, with focal lengths of 120 cm and 62 cm respectively. As the solar primary image and the cone are at the object focus of the first lens, the filter will therefore be crossed by a beam of parallel rays; the second lens will form an image of the corona at its focus F (figure 20), 62 cm away. The focal length ratio of the two lenses is 0.5 (the optical magnification), so that the equivalent focal length becomes 300 x 0.5 = 150 cm, which means that the Sun has a diameter of 1.4 cm at the output.

The elements of the coronagraph were located, along the z-optical axis, at the positions:

z = 0 cm, entrance objective, single lens Ø 25 cm, f = 300 cm

43 cm, secondary focus (two internal reflections), image Ø 0.4 cm

300 cm, primary focus, image Ø 2.8 cm, occulting cone, and field lens f = 85 cm

300 + 118 cm, pupil plane Ø 10 cm, Lyot diaphragm Ø 6.4 cm, Lyot spot Ø 0.2 cm

300 + 120 cm, 1st lens of the afocal system (γ = 0.5), f = 120 cm

300 + 125 cm, narrow band interference filter

300 + 130 cm, 2nd lens of the afocal system (γ = 0.5), f = 62 cm

300 + 130 + 62 cm, output image focus, equivalent focal length = 150 cm, image Ø 1.4 cm

At the output focus of figure 20, an electronic camera of Lallemand-type could be installed (figure 22), or a conventional photographic plate, or a small spectrograph. It was the spectrograph designed for the 1973 eclipse in Mauritania that was re-used. It had a magnification of 2/3, a grating of 10 x 10 cm², 600 grooves/mm, providing the dispersion of 12 Å/mm in the second order, and a spectral resolution of 0.3 Å. The useful spectral domain was the 3000 Å to 11000 Å waveband. The final detector was in all cases either a standard or nuclear (for the electronic camera) plate. The development of the plates was locally done; the digitization was performed after the missions, with a micro densitometer (such as the Joyce at Meudon or the PDS-Perkin Elmer of Institut d'Optique at Orsay), allowing to measure the photographic densities and convert them into numerical values that can be processed by a computer (at that time, the IBM 360/65 of INAG was available in Meudon, together with a PDP 11/34 running the software of the "Centre de Dépouillement des Clichés Astronomiques", the CDCA system developed at Nice by Albert Bijaoui ; a VAX11/780 came in 1981).

Figure 23 shows the coronagraph being tested in Meudon before its transport to Saint Véran in October 1975, as shown on figure 24.

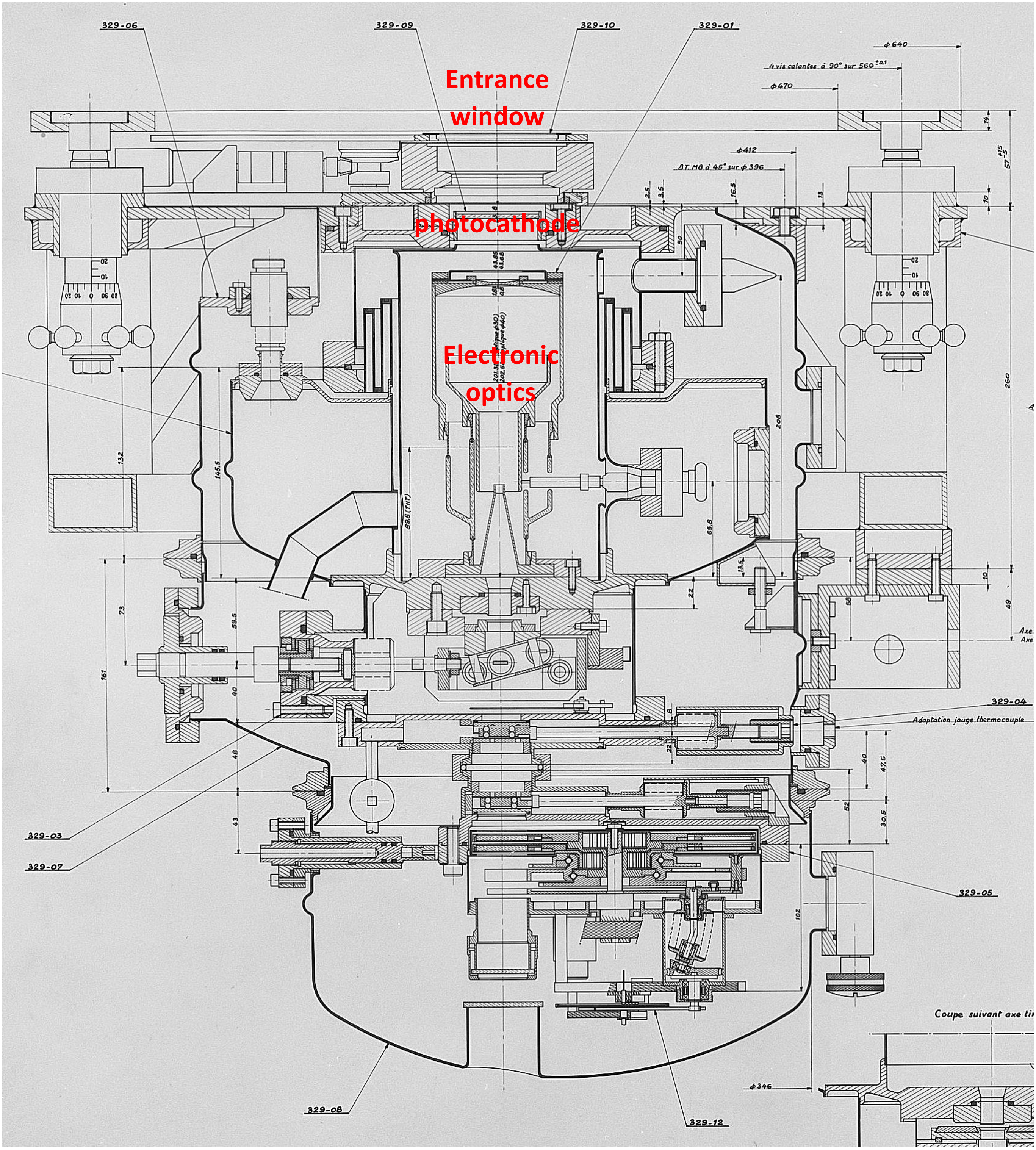

*Figure 22: cross section of the electronic valve camera showing its complexity (Georges Darré drawing, 1982, courtesy Paris observatory).*

The scientific program was based on the extreme sensitivity of the electronic valve camera, a version of the Lallemand detector (figure 22) to explore the hot corona at large distance. It allowed the observation of monochromatic images in the green and red lines (5303 Å of FeXIV, 6374 Å of FeX, Figure 25) up to 0.4 solar radius from the limb, and even 0.8 radius in the IR lines of FeXIII. Such observations make it possible to determine the electron density and temperature, and trace the direction of the coronal magnetic field (with linear polarization analysis). More specifically, the experiments envisaged were the following:

- Determine the electron density and temperature using neighboring ions, such as FeXIII and FeXIV.

- Plot the direction of the magnetic field by measuring the linear polarization of the IR line of FeXIII at 10747 Å and the green line of FeXIV at 5303 Å.
- Study the protuberance/corona interface (condensation, plasma transfer) using the Hα line (cold material at $10^4$ K) and the red line of FeX at 6374 Å at $10^6$ K (Figure 25).
- Study the oscillations of the prominences, provided that sufficient temporal resolution is available.
- Measure the magnetic field of the prominences in the IR line of HeI at 10830 Å, by measuring the Hanle linear depolarization and rotation.

It should be emphasized that the interest of the electronic camera was the sensitivity, linearity and absence of threshold, allowing, in comparison to the conventional photo plate, either to detect fainter coronal lines in spectroscopic mode (Figure 26) or to see higher in the corona with filters (provided that the sky permitted it). These advantages were offered a few years later by CCDs, without the disadvantages of vacuum and cryogenics.

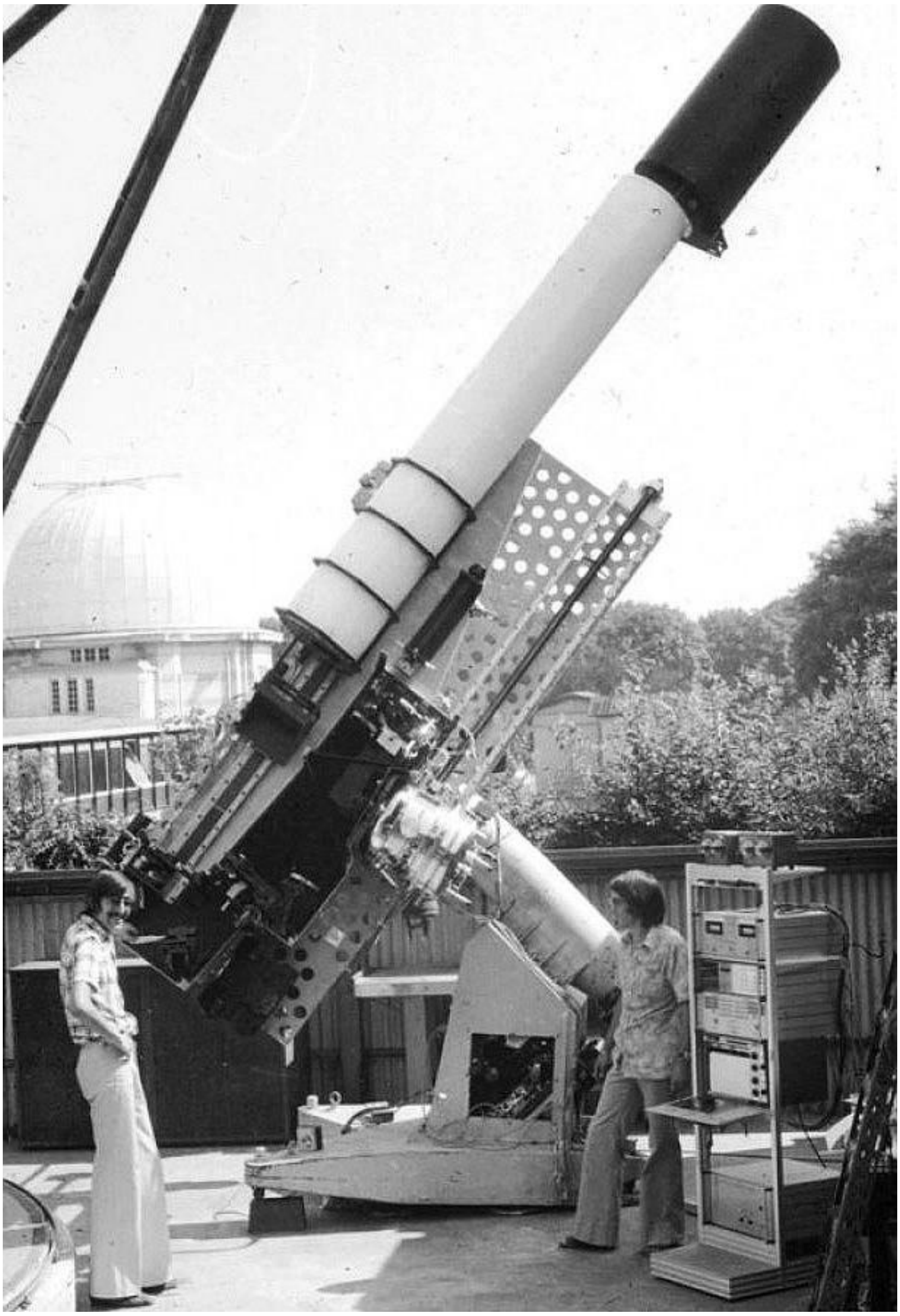

***Figure 23**: the coronagraph with the electronic valve camera, tested in Meudon (courtesy Paris observatory).*

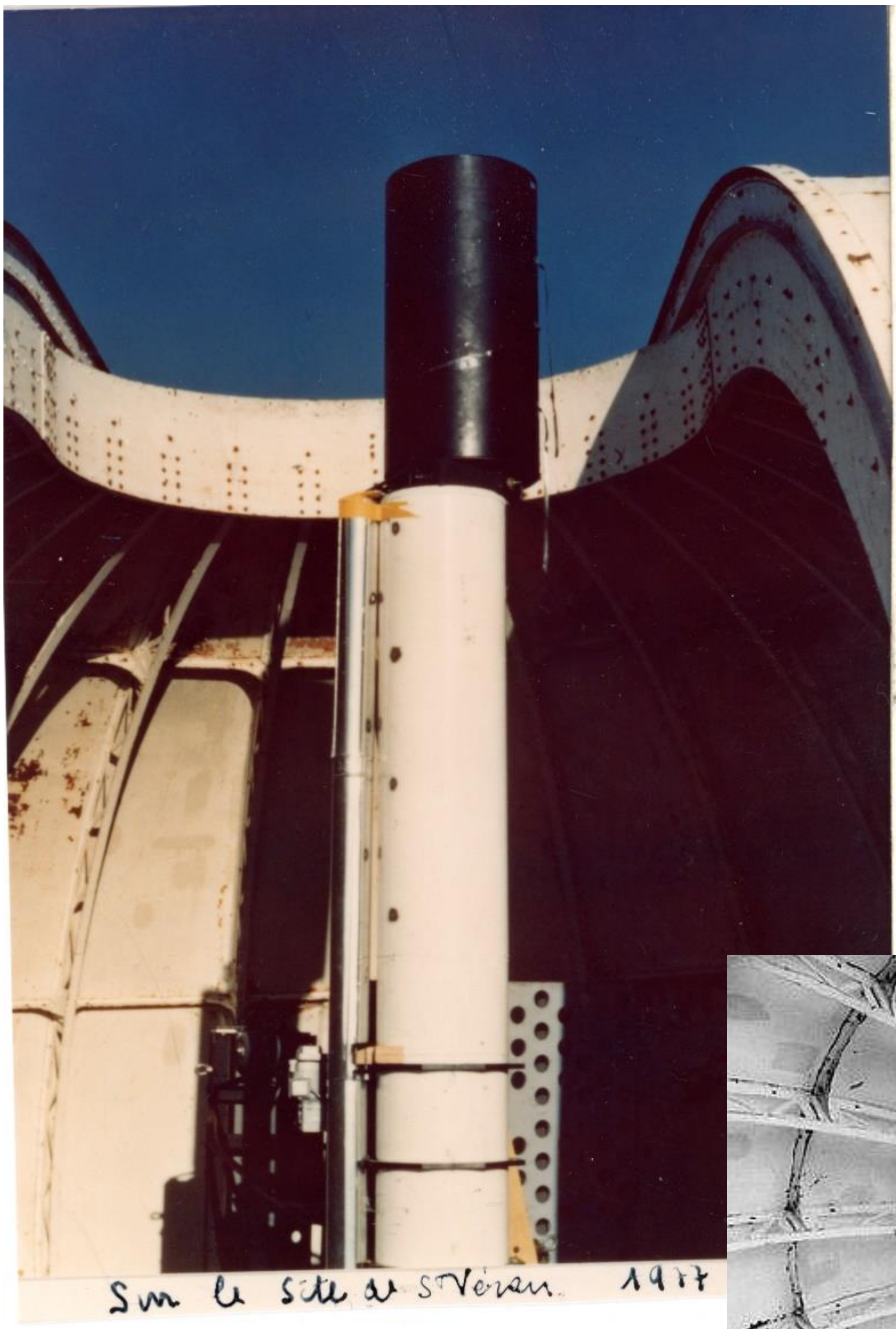

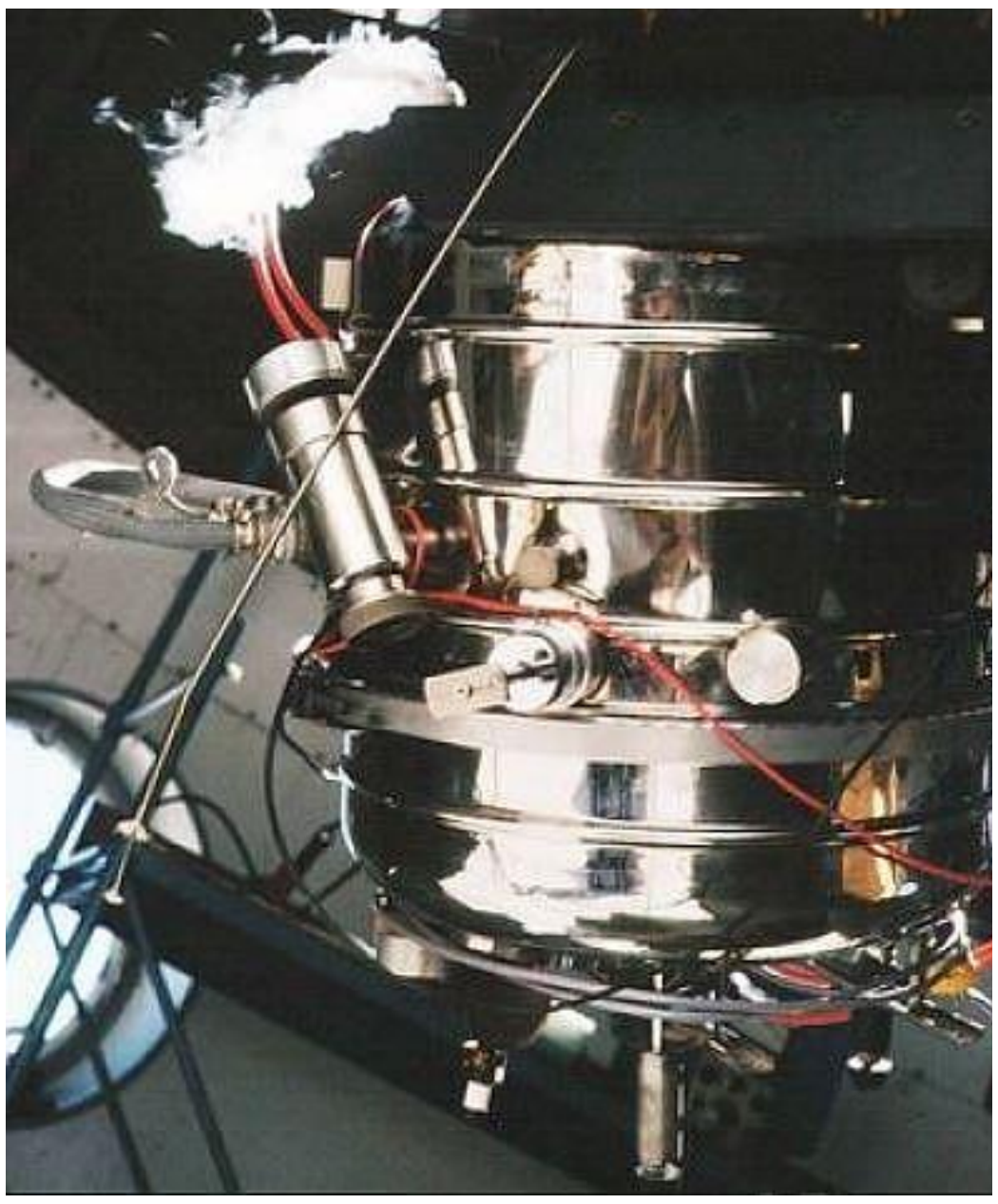

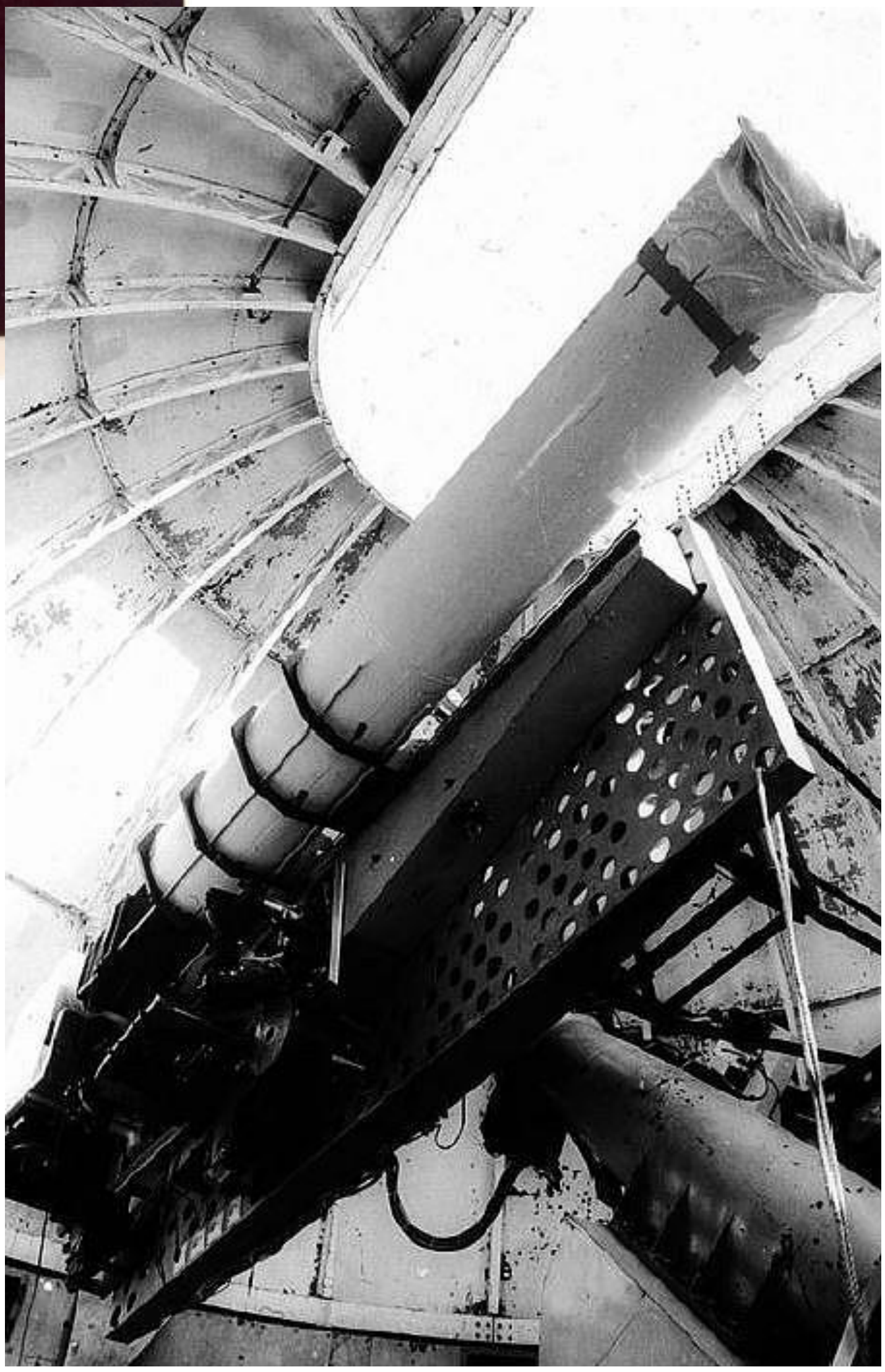

*Figure 24: the coronagraph in the dome at Saint Véran, in 1977, with the electronic camera (top right). Courtesy Paris observatory.*

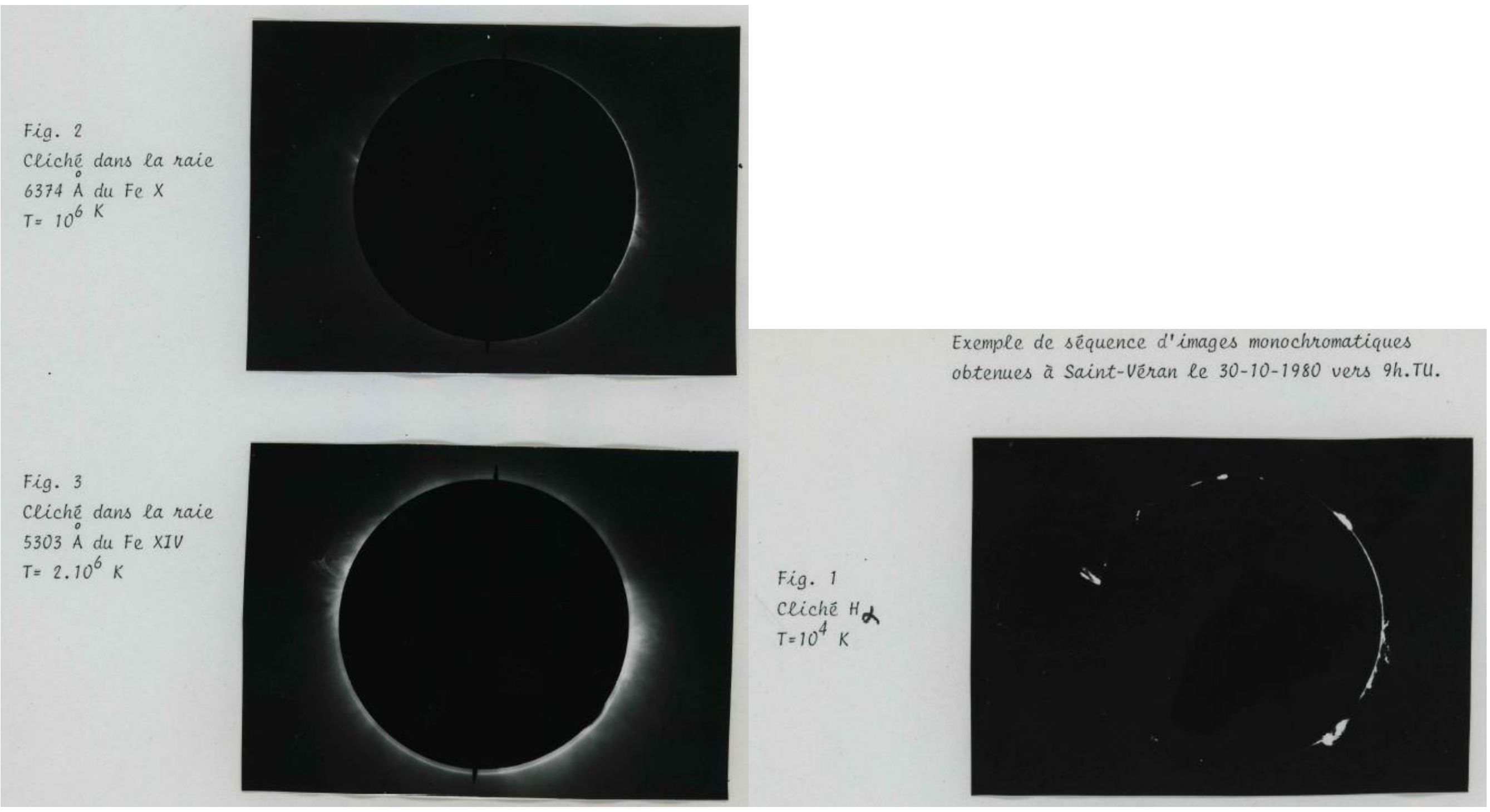

*Figure 25: monochromatic images of the corona and prominences with narrow bandpass filters, obtained in classic photography with the coronagraph of Saint Véran (courtesy Paris observatory). Left: hot corona, Fe X 6374 Å, $10^6$ K (red line) and Fe XIV 5303 Å, 2 $10^6$ K (green line). Right: cold corona, Hα 6563 Å, $10^4$ K (prominences).*

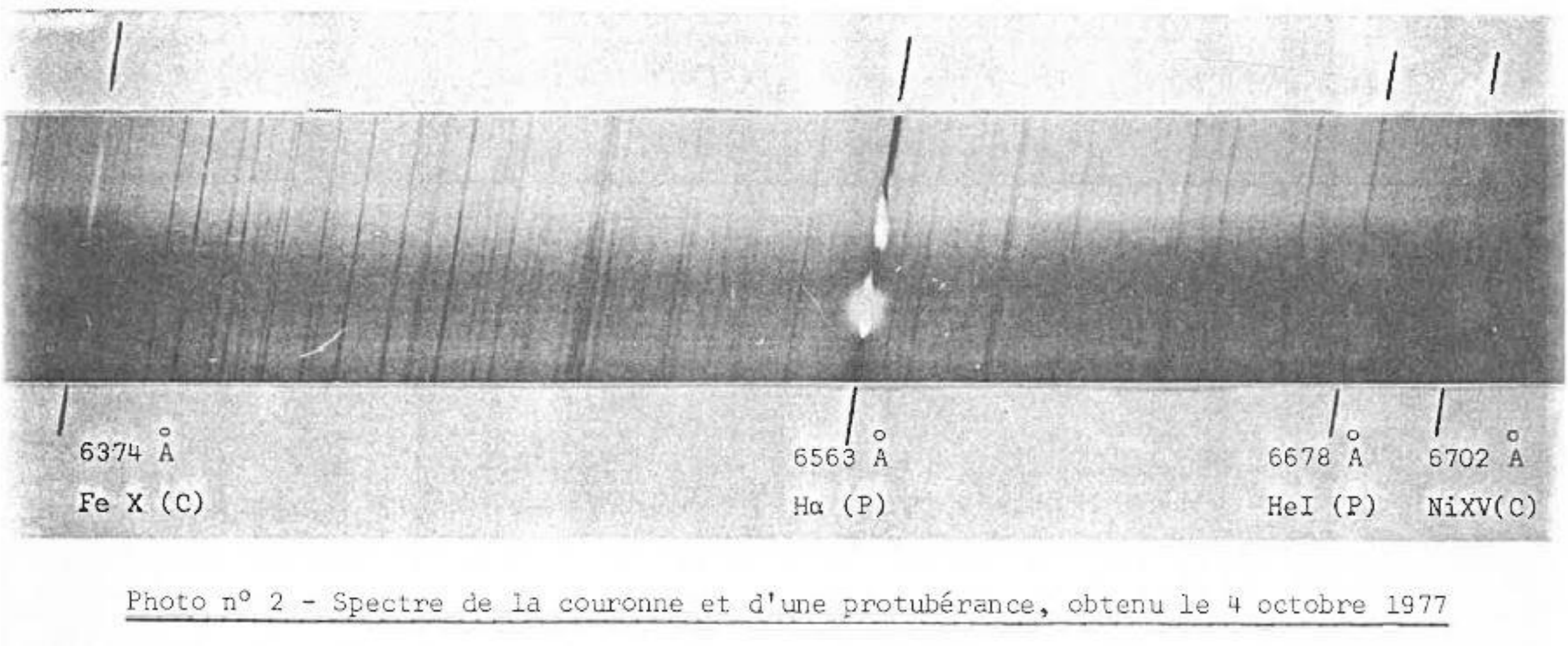

*Figure 26: spectrum of the hot corona (C) and prominences (P) obtained in classic photography at Saint Véran. The red line of iron (Fe X) corresponds to T = $10^6$ K, that of Ni XV to 2.3 $10^6$ K. The prominences, visible in Hα, are 100 times cooler ($10^4$ K) and much more bright than the hot corona (courtesy Paris observatory).*

## WHY THE SAINT VERAN OPERATION WAS NOT A FULL SCIENFIFIC SUCCESS

Two young scientific researchers (retired today) of Paris observatory (CNRS) used the coronagraph, Jean-Pierre Picat and Françoise Crifo, under the supervision of Paul Felenbok. It was certainly a very difficult task for them, because we did not find, unfortunately, any scientific publication with Saint Véran observations. Why ? The electronic camera met some problems, and it was necessary to return to a classic 35 mm reflex camera with a motorized magazine of 250 frames to produce the few spectra and coronal images of figures 25 and 26, showing that the coronagraph worked perfectly. But observations with such a detector were rather limited to the low corona.

We saw that the coronagraph project at Saint Véran was in the continuity of the total eclipse missions (1970, 1973) with electronic cameras. During an eclipse, the sky is black, and it is possible to observe the high corona at several solar radii. This was the purpose of the electronic camera, which was created for night time observations of deep-sky objects. Trying to detect coronal structures far from the Sun, in the forbidden lines of iron, with such a device, 100 times more sensitive than the photo plate (almost a modern CCD), was a truly original and audacious goal. My personal opinion is that the choice of the Lallemand camera, a fantastic but complex tool with constraints of vacuum and cooling, was probably too ambitious for Saint Véran, because of the lack of logistics, human and financial resources. This was especially the case, when one has to ensure the start-up of a new high-altitude station with complicated access. Rozelot already mentioned these difficulties, even at the Pic du Midi, in 1972 (where access conditions were much better with a cable car). The classic photography, introduced later (1979), saved the project, but it was no longer as innovative compared to what was done elsewhere, so that it became difficult to produce new and original results of international level.

We can also notice the growing competition from space, at first instruments using 35 mm films, such as the white light coronagraph onboard SKYLAB/NASA (Apollo Telescope Mount platform) from 1973 to 1979 (figure 27, see MacQueen *et al*, 1974), then those equipped with vidicon tubes, such as the SMM coronagraph, with real time digitization (Figure 28, see MacQueen *et al*, 1980). NASA's Solar Maximum Mission satellite only operated for six months in 1980 (March-September) and unfortunately failed. It was later repaired in April 1984 by an extravehicular spacewalk from the American space shuttle, when the coronagraph of Saint Véran was already closed, so there were only 6 months of intersection between them. The SMM coronagraph then operated until 1989 with filters and polarimetry. Later, LASCO's wide-field coronagraphs onboard SOHO used CCDs (Figure 29, up to 30 solar radii, working since 1996) and attested the importance of coronal observation, because in 2024 they are still in operation by ESA and NASA; this is also the case for the EUV telescopes of the EIT/SOHO instrument, which produces images of the lower corona in the hot lines of highly ionized iron (but in this wavelength range, Solar Dynamics Observatory SDO/NASA does much better since 2010).

The start-up of the CFHT in Hawaii (1979) undoubtedly precipitated the rapid end of the Saint Véran operation, as the team published his first results with the CFHT and the electronic camera as early as 1982 (Baudrand *et al*, 1982; Picat *et al*, 1982).

The coronagraph lasted only a few years, as the conditions met at 2930 m to operate a sophisticated instrument were too difficult. The future of the 7-metre dome was then questioned, and in the expert report (1981) of G. Rousset, we read:

"A new dismantling and reassembly of this dome, if not impossible, presents great difficulties. Indeed, such an operation cannot be conceived without:

- completely review the rotational motion path and its driving system
- completely overhaul the trapdoors opening and closing system
- redo the polyurethane spray that would be fatally destroyed by the dismantling, and the painting

Because of these major problems, I think that such a project could not provide a serious, fast and economical means available to researchers".

As a result, the coronagraph was removed, but the dome and the living base remained on the site, which stayed unoccupied for several years.

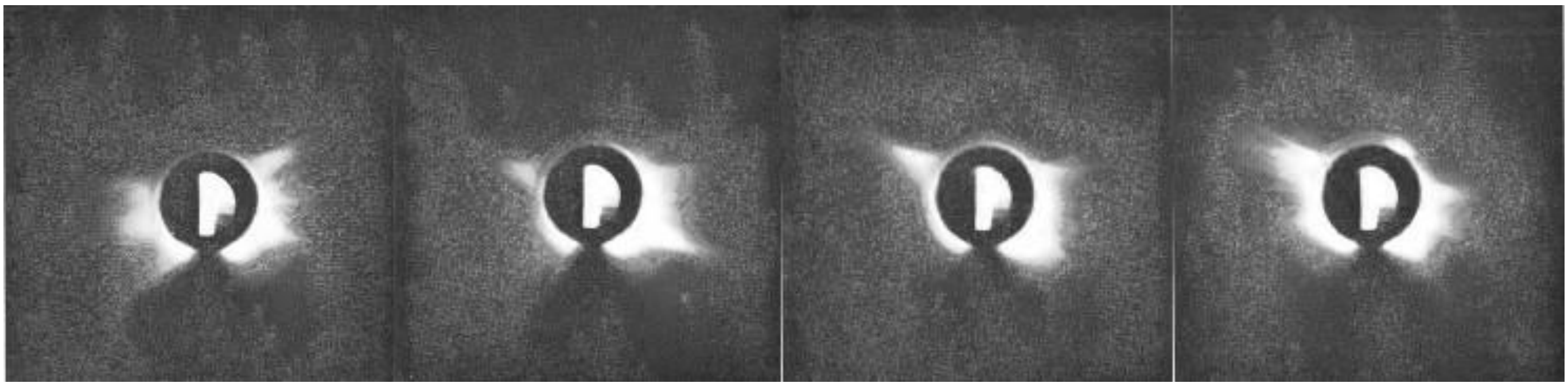

***Figure 27**: A space competitor, SKYLAB (NASA), here white light images of the corona on 28 May, 1, 5, 10 June 1973 (after MacQueen et al, 1974)*

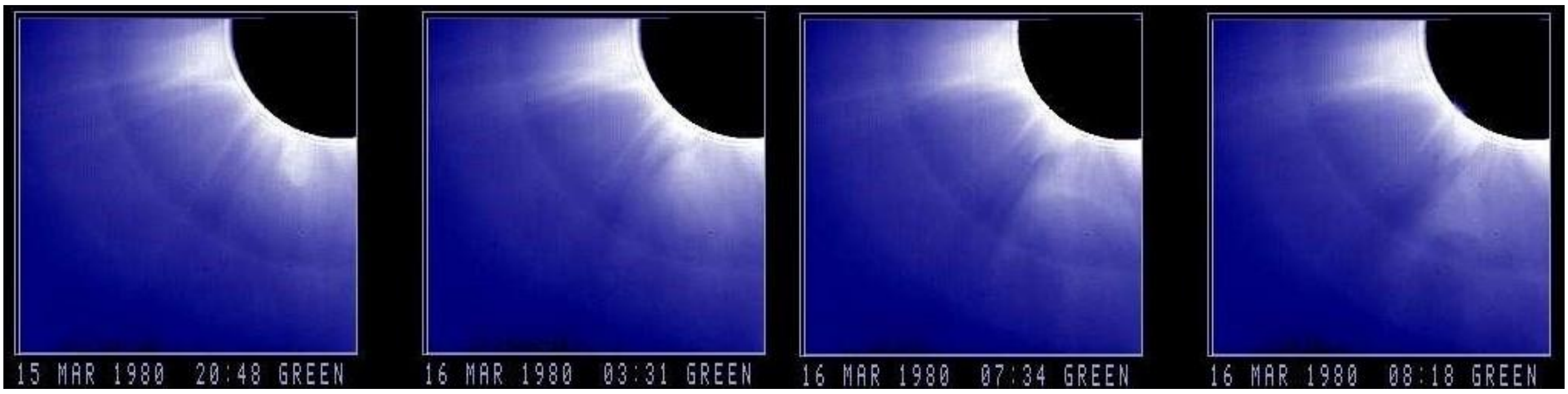

***Figure 28***: *A space competitor for 6 months in 1980, Solar Maximum Mission, here the coronal mass ejection of 15 March 1980 (courtesy NASA and High Altitude Observatory)*

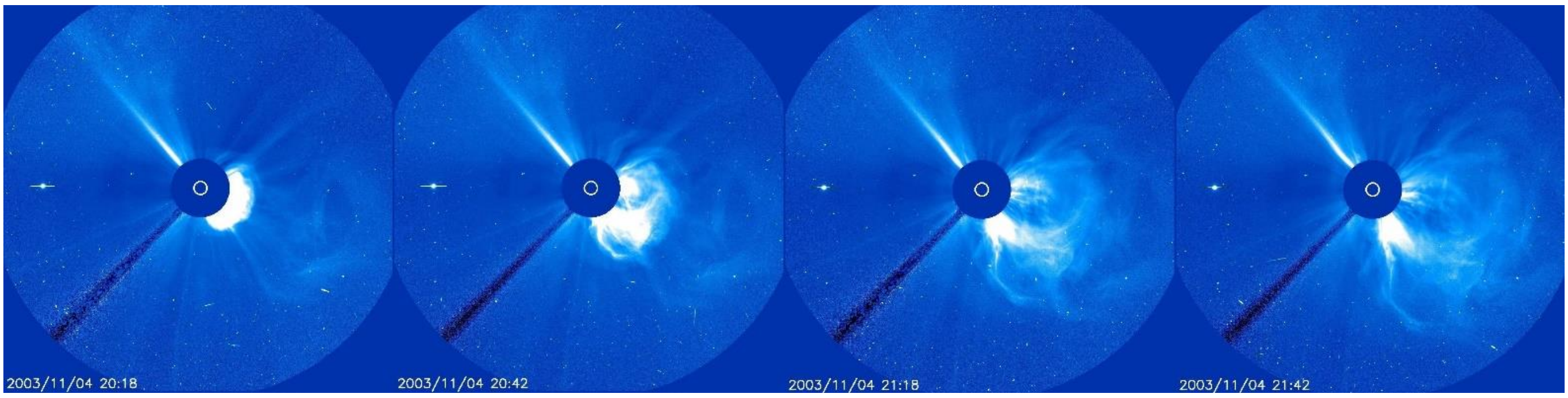

***Figure 29***: *C3 SOHO/LASCO wide-field coronagraph running after 1995, here the coronal mass ejection of 4 November 2003 (courtesy ESA/NASA)*

**TOWARDS A SECOND LIFE FOR THE OBSERVATORY**

The second life of the Saint Véran station began after a visit by Paul Felenbok and Jacques Léorat in the summer of 1988: they were surprised to see the good conservation of the facilities after 6 years of closure; Léorat then tried to interest amateur astronomers and promote the dissemination of scientific culture, with the agreement of Paris Observatory. The next step was in 1989, when the AstroQueyras association of amateur astronomers was created and took in charge the observatory. They installed in the 7-metre dome a 62 cm Cassegrain telescope (Figure 30) on loan from Observatoire de Haute Provence (OHP), which was open to amateurs in 1990.

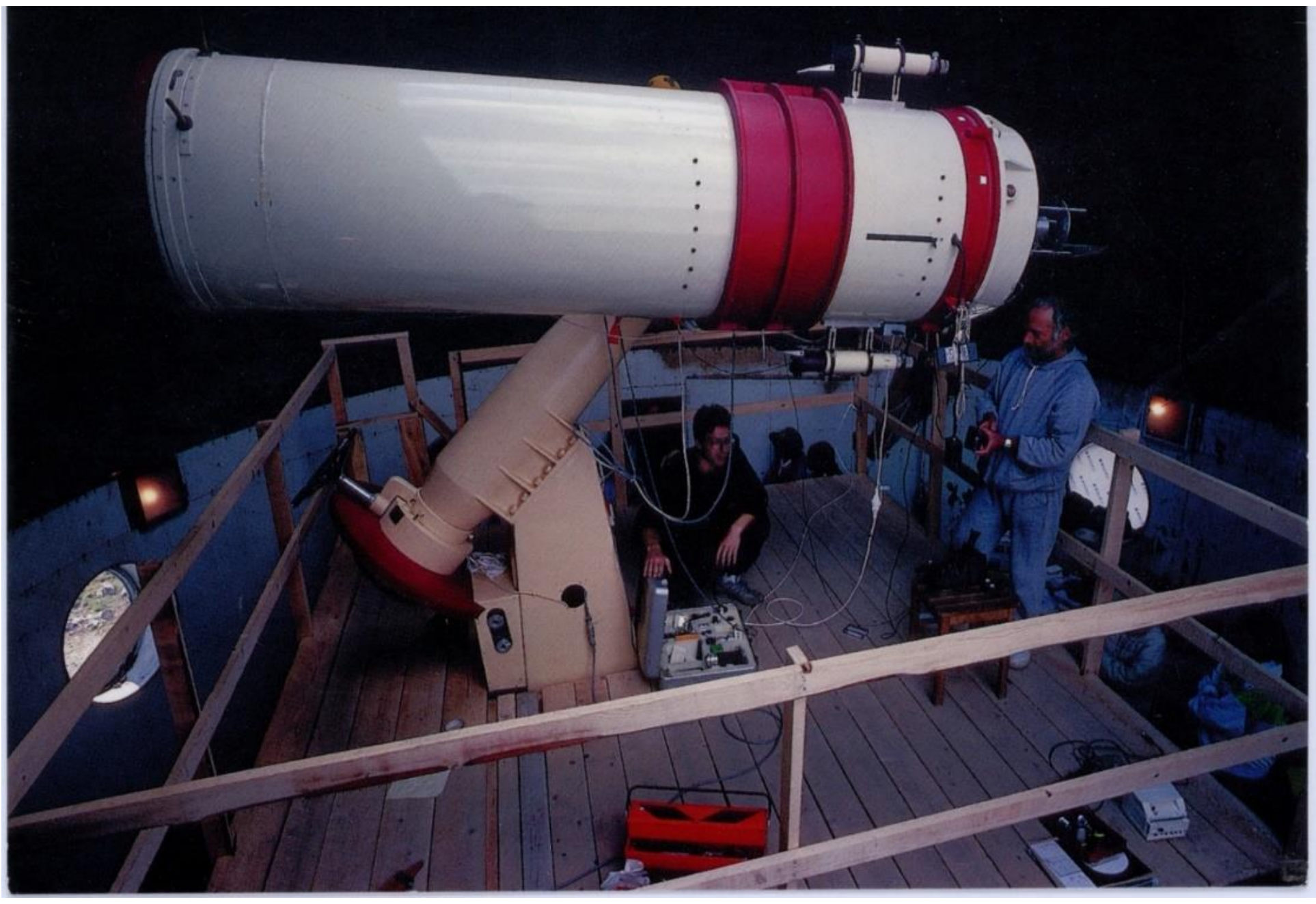

***Figure 30***: *the 62 cm telescope in the 7-metre dome (courtesy Paris observatory)*

The observatory developed progressively with the adjunction of two domes, both with 50 cm aperture night telescopes. In 2015, the capacity of the station was extended from 6 to 18 visitors and the living areas were totally rebuilt (figure 31). Solar photovoltaic panels provide now electric energy. In parallel, "la Maison du Soleil' (the Sun's house), in Saint Véran village, was created in 2016, in collaboration with Paris observatory. It is a modern interpretation space for the public devoted to the nature of the Sun. It includes professional instruments, operated by the staff: a coelostat, a 30 cm solar refractor to produce a large white light image, and a 7-m high resolution spectrograph fed by a 50 cm telescope, to show the Doppler effect, all provided by Paris-Meudon observatory.

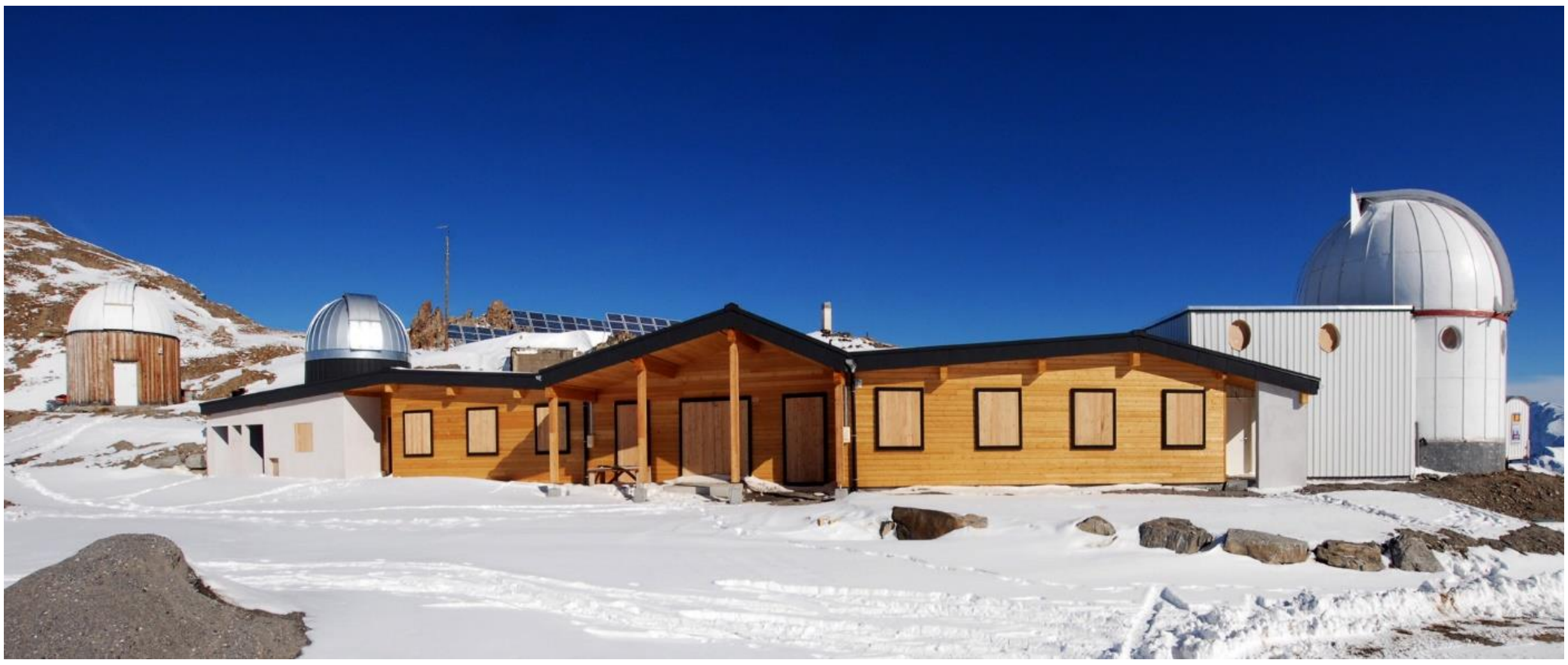

***Figure 31****: the observatory after reconstruction in 2015 (courtesy D. Menel)*

## CONCLUSION

The foundation of the high altitude (2930 m) astronomical station at Saint Véran was a fantastic and difficult adventure led by Pauf Felenbok. Everything had to be constructed, such as a track, a living house, a dome, or to be brought, such as electricity and heating generators; the project was complicated by the altitude and the lack of human and financial resources. The operating of the electronic camera (need of vacuum and cooling at 77 K) was probably the most difficult task at Saint Véran, the pressure to use the new CFHT telescope was increasing, so that the station stopped in 1982. However, despite the lack of major scientific results, it is clear, forty years later, that it was a beautiful technical success. Professional astronomers proposed later other projects of night telescopes, such as the large 3-metre polar telescope (POST, 1990), or recently a 1-metre robotic telescope, which were unsuccessful because of the complicated logistics. Another robotic solar telescope (MeteoSpace) was also proposed in 2015 and finally built at Calern plateau (Côte d'Azur observatory). The problem with robotic telescope, which operate remotely or automatically, is again the access and the lack of permanent staff in case of failure. But amateur astronomers succeeded to develop the observatory after 1989, with three night telescopes of the 50 cm class; the accommodation capacity passed in 2015 from 6 to 18 visitors, allowing the public to come for one night; a Sun's house was created in Saint Véran village in close collaboration with Paris observatory. Paul Felenbok is the father of all these successful actions, that's why the observatory honours his name.

## ACKNOWLEDGEMENTS

The author thanks Dr Jacques Léorat for fruitful suggestions. He is indebted to Isabelle Bualé for recovering and digitizing Paul Felenbok's archives concerning Saint Véran.

**Web sites**